\documentclass[11pt,a4paper]{article}

\usepackage[hidelinks]{hyperref}
\usepackage{amssymb, amsmath, graphicx, subcaption}
\usepackage{amsthm}
\usepackage{autobreak}
\usepackage[ruled]{algorithm2e}
\usepackage{multirow}
\usepackage{float}
\usepackage{fancyhdr}

\usepackage{geometry}
\fancypagestyle{firststyle}
{
	\fancyhf{}
	\fancyhead[RE,LO]{}
	\fancyfoot[RE,LO]{}
	\fancyfoot[LE,RO]{Page \thepage}
}

\begin{document}
	
	\title{Analytic frozen orbits under the zonal harmonics perturbation from an Earth-like planet}
	
	\author{David Arnas\thanks{Purdue University, West Lafayette, IN 47907, USA. Email: \textsc{darnas@purdue.edu}}}
	
	\date{}	
	
	\maketitle
	
	\thispagestyle{firststyle}
	
	\begin{abstract}
		This work focuses on providing closed form analytical expressions to define frozen orbits under the effects of the zonal harmonics of an Earth-like planet. Particularly, the perturbation effects from the terms $J_2$, $J_3$, $J_4$, $J_5$, $J_6$, and $J_7$ are considered in this work. This is done using a power series expansion in the small parameter that allow not only to provide an approximate solution to the system, but also to determine the analytical expressions that define the initial osculating conditions that generate frozen orbits. As a result of that, the proposed methodology allows to study the bifurcation of frozen orbits close to the critical inclination by purely analytical methods. Additionally, the derivation of the secular variation of the orbital elements as well as the transformation from osculating to mean elements is provided based on the second order analytical solution proposed in this work. Examples of application are also provided to show the error performance of the results included in this document.
	\end{abstract}

\section{Introduction}

Frozen orbits are by far the most important set of orbits is aerospace engineering. This is due to their periodic properties as well as their long-term stability. Because of that, frozen orbits have been extensively used in a large number of space missions. For instance, Earth observation missions require in the majority of cases to use frozen orbits in order to maintain the same observation conditions when targeting a specific region on the Earth's surface. This is achieved by preventing the eccentricity vector to have a secular variation, which is one of the properties of frozen orbits. However, there are not the only case of application: missions for telecommunications and global and regional coverage also benefit from these kind of orbits as they provide periodicity to their dynamics. 

In general, frozen orbits can be defined either with analytical, numerical or semi-analytical methods, having each one of these approaches their advantages and disadvantages. For specific orbits, it is relatively simple to create a numerically iterative method to search for these frozen orbits. However, due to the nature of these methods, these results are difficult to extrapolate to other orbits and provide little insight on the problem itself. Conversely, analytical methods, although more complex to generate, can provide more general results applicable to a wider set of problems and, more importantly, a better insight on the dynamics of these systems. This is, for instance, one of the reasons why analytical approaches are still extensively used when performing mission design.

As a result of that, a large number of perturbation methods have been proposed over the years to study frozen orbits. This started with the study of the general main satellite problem (when only the perturbation effect of the oblatness of the main celestial body is considered) with the solutions of Brouwer~\cite{brouwer}, Kozai~\cite{kozai1959motion,kozai1962second}, Hori~\cite{hori}, or Deprit~\cite{deprit1969}, and the specific study of frozen orbits under this perturbation~\cite{coffey1986critical}. Later, these solutions were complemented and extended to include other terms of the zonal harmonics~\cite{coffey_zonal,liu_zonal,frozen_moon}, and the sectorial and tesseral terms of the gravitational field~\cite{wnuk,lara_tesseral}. 

In this work, however, we focus on the zonal harmonics problem, specifically, when the terms $J_2$, $J_3$, $J_4$, $J_5$, $J_6$, and $J_7$ from the gravitational field are considered. Additionally, this work assumes that all the terms of the zonal harmonics (apart from $J_2$) have an order of magnitude similar to $J_2^2$. This corresponds with the motion about a main celestial body similar in mass distribution to Earth. The objective of this work then is to first provide a second order analytical approximation to this problem. Second, to identify the analytical transformation from osculating to mean elements of the solution. Third, to determine the secular variation of the orbital elements. And finally, to use the previous results to identify and study frozen orbits. This includes providing closed-form expressions to study the bifurcation that appears close to the critical inclination. These results are provided using a simple perturbation approach based on a power series expansion on the small parameter and using a complete osculating formulation as done in Arnas~\cite{meanj2,frozenj2}.

Compared with other approaches, the proposed solution does not rely on any kind of averaging technique as in other works~\cite{kamel1969expansion,kamel1970perturbation,deprit1970main,deprit1981elimination,deprit1982delaunay,cid,abad2001short,lara2014delaunay,mahajan2018exact,lara2019new,abad2020integration,abad2021cid}. This allows to directly analyze the problem from an osculating perspective, providing additional insight into the system. An example of that is the possibility to derive closed-form expressions that define the frozen orbits close to the critical inclination, a problem that had to be solved by numerical continuation in previous works both for the $J_2$ problem~\cite{coffey1986critical,Broucke}, and the general zonal harmonics problem~\cite{elipe}. Additionally, and compared with other perturbation methods based on osculating elements~\cite{zonal,koopman,schur}, this approach provides a more simple process to study frozen orbits with good accuracy. This allows to easily obtain second order solutions as well as to study the bifurcation close to the critical inclination.

This paper is organized as follows. First, the dynamical system is presented as well as the variable transformations used in this work. This includes the definition of the orbital elements, as well as the time regularization used. Second, the perturbation method to study near circular frozen orbits is presented and applied to the terms of the zonal harmonics considered. Third, the approximate analytical solution of this approach is included. Fourth, the transformation from osculating to mean elements and the secular variation of the orbital elements is derived based on the previous result. Fifth, the frozen conditions are derived both for low eccentric orbits and for eccentric orbits close to the critical inclination. In here, closed-form expressions to study bifurcation are included. Finally, some examples of application are included to show the performance of the analytical methods presented in this work.


\section{Dynamical model}

This work focuses on the study of the effects of the perturbation produced by the zonal harmonics terms of the gravitational potential of an Earth-like planet, and more specifically, we will center our attention in the terms $J_2$, $J_3$, $J_4$, $J_5$, $J_6$, and $J_7$ of the potential. Particularly, their effects can be represented by the following Hamiltonian:
\begin{eqnarray}
	\mathcal{H} & = & \displaystyle\frac{1}{2}\left(p_r^2 + \frac{p_{\varphi}^2}{r^2} + \frac{p_{\lambda}^2}{r^2\cos^2(\varphi)}\right) - \frac{\mu}{r} + \sum_{n=2}^{m} \mu J_n \frac{R^n}{r^{n+1}} P_n(\sin(\varphi)),
\end{eqnarray}
where:
\begin{eqnarray}
	p_r & = & \dot{r}; \nonumber \\
	p_{\varphi} & = & r^2\dot{\varphi}; \nonumber \\
	p_{\lambda} & = & r^2\cos^2(\varphi)\dot{\lambda};
\end{eqnarray}
represent the conjugate momenta of the radial distance $r$, the latitude of the orbit $\varphi$, and the inertial longitude of the orbit $\lambda$, respectively. Additionally, $\mu$ and $R$ are the gravitational constant and the mean equatorial radius of the main celestial body, $P_n(x)$ are the set of Legendre polynomials of order $n$ in variable $x$, and $m$ is the maximum order of the zonal harmonics included in the model. From this Hamiltonian, it is possible to obtain Hamilton's equations in spherical coordinates:
\begin{eqnarray} \label{eq:hamilton}
\displaystyle\frac{dr}{dt} & = & p_r; \nonumber \\
\displaystyle\frac{dp_r}{dt} & = & -\frac{\mu}{r^2} + \frac{p_{\varphi}^2}{r^3}  + \frac{p_{\lambda}^2}{r^3\cos^2(\varphi)} + \sum_{n=2}^m (n+1)\mu J_n P_n(\sin(\varphi))\frac{R^n}{r^{n+2}}; \nonumber \\
\displaystyle\frac{d\varphi}{dt} & = & \frac{p_{\varphi}}{r^2}; \nonumber \\
\displaystyle\frac{dp_{\varphi}}{dt} & = & -\frac{p_{\lambda}^2}{r^2}\frac{\sin(\varphi)}{\cos^3(\varphi)} - \sum_{n=2}^m \mu J_n \frac{\partial P_n(\sin(\varphi))}{\partial\varphi}\frac{R^n}{r^{n+1}}; \nonumber \\
\displaystyle\frac{d\lambda}{dt} & = & \frac{p_{\lambda}}{r^2\cos^2(\varphi)}; \nonumber \\
\displaystyle\frac{dp_{\lambda}}{dt} & = & 0.
\end{eqnarray}

Once Hamilton equations are obtained, the objective is to perform a series of variable transformations and a time regularization in order to have a differential equation in which to apply the same perturbation method seen in Arnas~\cite{frozenj2}.

\subsection{Variable transformation}

Instead of using the common keplerian orbital elements: semi-major axis ($a$), eccentricity ($e$), inclination ($i$), argument of perigee ($\omega$), right ascension of the ascending node ($\Omega$), and true anomaly $\nu$, this work makes use of the orbital elements used in Arnas~\cite{meanj2}, namely: $\{A, e_x, e_y, i, \Omega, \theta\}$, where A is defined as:
\begin{equation}
    A = \left(\displaystyle\frac{R}{a(1-e^2)}\right)^2 = \left(\displaystyle\frac{\mu R \cos^2(\varphi)}{p_{\varphi}^2\cos^2(\varphi) + p_{\lambda}^2}\right)^2.
\end{equation}
$e_x$ and $e_y$ are the two components of the eccentricity vector in the orbital plane:
\begin{eqnarray}
    e_x & = & e \cos(\omega), \nonumber \\
    e_y & = & e \sin(\omega),
\end{eqnarray}
and $\theta = \omega + \nu$ is the argument of latitude of the orbiting object. This transforms Eq.~\eqref{eq:hamilton} into the following system of differential equations:
\begin{eqnarray} \label{eq:difftheta}
	\displaystyle\frac{dA}{dt} & = & \sqrt[\leftroot{-1}\uproot{2}\scriptstyle 4]{\frac{\mu^2 A^3}{R^6}} \sum_{n=2}^{m} 4 J_n \frac{\partial P_n(\sin(\theta)\sin(i))}{\partial (\sin(\theta)\sin(i))} \left(\sqrt{A}\right)^{n+2} \nonumber \\
	& \times & \sin(i)\cos(\theta)(1 + e_x\cos(\theta) + e_y\sin(\theta))^{n+1}; \nonumber \\
	\displaystyle\frac{de_x}{dt} & = & \sqrt[\leftroot{-1}\uproot{2}\scriptstyle 4]{\frac{\mu^2 A^3}{R^6}} \Big[-\sum_{n=2}^{m} J_n \frac{\partial P_n(\sin(\theta)\sin(i))}{\partial (\sin(\theta)\sin(i))} \left(\sqrt{A}\right)^{n}(1 + e_x\cos(\theta) + e_y\sin(\theta))^{n+1} \nonumber \\
	& \times & \Big(e_y \displaystyle\frac{\cos^2(i)}{\sin(i)}\sin(\theta) +  \sin(i)\cos(\theta)(2\cos(\theta) + (1 + \cos^2(\theta))e_x + \sin(\theta)\cos(\theta)e_y)\Big) \nonumber \\
	& + & \sum_{n=2}^{m} J_n P_n(\sin(\theta)\sin(i))\left(\sqrt{A}\right)^{n} \sin(\theta)(1 + e_x\cos(\theta) + e_y\sin(\theta))^{n+2}\Big]; \nonumber \\
	\displaystyle\frac{de_y}{dt} & = & \sqrt[\leftroot{-1}\uproot{2}\scriptstyle 4]{\frac{\mu^2 A^3}{R^6}} \Big[\sum_{n=2}^{m} J_n \frac{\partial P_n(\sin(\theta)\sin(i))}{\partial (\sin(\theta)\sin(i))} \left(\sqrt{A}\right)^{n}(1 + e_x\cos(\theta) + e_y\sin(\theta))^{n+1} \nonumber \\
	& \times & \Big(e_x \displaystyle\frac{\cos^2(i)}{\sin(i)}\sin(\theta) -  \sin(i)\cos(\theta)(2\sin(\theta) + \sin(\theta)\cos(\theta)e_x + (1 + \sin^2(\theta))e_y)\Big) \nonumber \\
	& - & \sum_{n=2}^{m} J_n P_n(\sin(\theta)\sin(i))\left(\sqrt{A}\right)^{n} \cos(\theta)(1 + e_x\cos(\theta) + e_y\sin(\theta))^{n+2}\Big]; \nonumber \\
	\displaystyle\frac{di}{dt} & = & -\sqrt[\leftroot{-1}\uproot{2}\scriptstyle 4]{\frac{\mu^2 A^3}{R^6}} \sum_{n=2}^{m} J_n \frac{\partial P_n(\sin(\theta)\sin(i))}{\partial (\sin(\theta)\sin(i))} \left(\sqrt{A}\right)^{n} \nonumber \\
	& \times & \cos(i)\cos(\theta)(1 + e_x\cos(\theta) + e_y\sin(\theta))^{n+1}; \nonumber \\
	\displaystyle\frac{d\Omega}{dt} & = & -\sqrt[\leftroot{-1}\uproot{2}\scriptstyle 4]{\frac{\mu^2 A^3}{R^6}} \sum_{n=2}^{m} J_n \frac{\partial P_n(\sin(\theta)\sin(i))}{\partial (\sin(\theta)\sin(i))} \left(\sqrt{A}\right)^{n} \nonumber \\
	& \times & \displaystyle\frac{\cos(i)}{\sin(i)}\sin(\theta)(1 + e_x\cos(\theta) + e_y\sin(\theta))^{n+1}; \nonumber \\
	\displaystyle\frac{d\theta}{dt} & = &  \sqrt[\leftroot{-1}\uproot{2}\scriptstyle 4]{\frac{\mu^2 A^3}{R^6}} \big(1 + \sum_{n=2}^{m} J_n \frac{\partial P_n(\sin(\theta)\sin(i))}{\partial (\sin(\theta)\sin(i))} \left(\sqrt{A}\right)^{n} \nonumber \\
	& \times & \displaystyle\frac{\cos^2(i)}{\sin(i)}\sin(\theta)(1 + e_x\cos(\theta) + e_y\sin(\theta))^{n-1}\big)\left(1 + e_x\cos(\theta) + e_y\sin(\theta)\right)^2.
\end{eqnarray}

\subsection{Time regularization}

The goal of this step is to change the independent variable from the system of differential equations such that the evolution of the system is provided by the argument of latitude $\theta$. This means that we are effectively making the time evolution of the system to become a function of the argument of latitude as any other orbital element. Therefore, by doing this time regularization with $\theta$, the following system of differential equations is obtained:
\begin{eqnarray} \label{eq:difftheta}
	\displaystyle\frac{dA}{d\theta} & = & \frac{1}{\Delta} \sum_{n=2}^{m} 4 J_n \frac{\partial P_n(\sin(\theta)\sin(i))}{\partial (\sin(\theta)\sin(i))} \left(\sqrt{A}\right)^{n+2} \nonumber \\
	& \times & \sin(i)\cos(\theta)(1 + e_x\cos(\theta) + e_y\sin(\theta))^{n-1}; \nonumber \\
	\displaystyle\frac{de_x}{d\theta} & = &  \frac{1}{\Delta} \Big[-\sum_{n=2}^{m} J_n \frac{\partial P_n(\sin(\theta)\sin(i))}{\partial (\sin(\theta)\sin(i))} \left(\sqrt{A}\right)^{n}(1 + e_x\cos(\theta) + e_y\sin(\theta))^{n-1} \nonumber \\
	& \times & \Big(e_y \displaystyle\frac{\cos^2(i)}{\sin(i)}\sin(\theta) +  \sin(i)\cos(\theta)(2\cos(\theta) + (1 + \cos^2(\theta))e_x + \sin(\theta)\cos(\theta)e_y)\Big) \nonumber \\
	& + & \sum_{n=2}^{m} J_n P_n(\sin(\theta)\sin(i))\left(\sqrt{A}\right)^{n} \sin(\theta)(1 + e_x\cos(\theta) + e_y\sin(\theta))^{n}\Big]; \nonumber \\
	\displaystyle\frac{de_y}{d\theta} & = & \frac{1}{\Delta} \Big[\sum_{n=2}^{m} J_n \frac{\partial P_n(\sin(\theta)\sin(i))}{\partial (\sin(\theta)\sin(i))} \left(\sqrt{A}\right)^{n}(1 + e_x\cos(\theta) + e_y\sin(\theta))^{n-1} \nonumber \\
	& \times & \Big(e_x \displaystyle\frac{\cos^2(i)}{\sin(i)}\sin(\theta) -  \sin(i)\cos(\theta)(2\sin(\theta) + \sin(\theta)\cos(\theta)e_x + (1 + \sin^2(\theta))e_y)\Big) \nonumber \\
	& - & \sum_{n=2}^{m} J_n P_n(\sin(\theta)\sin(i))\left(\sqrt{A}\right)^{n} \cos(\theta)(1 + e_x\cos(\theta) + e_y\sin(\theta))^{n}\Big]; \nonumber \\
	\displaystyle\frac{di}{d\theta} & = & -\frac{1}{\Delta} \sum_{n=2}^{m} J_n \frac{\partial P_n(\sin(\theta)\sin(i))}{\partial (\sin(\theta)\sin(i))} \left(\sqrt{A}\right)^{n} \nonumber \\
	& \times & \cos(i)\cos(\theta)(1 + e_x\cos(\theta) + e_y\sin(\theta))^{n-1}; \nonumber \\
	\displaystyle\frac{d\Omega}{d\theta} & = & -\frac{1}{\Delta} \sum_{n=2}^{m} J_n \frac{\partial P_n(\sin(\theta)\sin(i))}{\partial (\sin(\theta)\sin(i))} \left(\sqrt{A}\right)^{n} \nonumber \\
	& \times & \displaystyle\frac{\cos(i)}{\sin(i)}\sin(\theta)(1 + e_x\cos(\theta) + e_y\sin(\theta))^{n-1}; \nonumber \\
	\displaystyle\frac{dt}{d\theta} & = & \sqrt[\leftroot{-1}\uproot{2}\scriptstyle 4]{\frac{R^6}{\mu^2 A^3}}\frac{1}{\Delta \left(1 + e_x\cos(\theta) + e_y\sin(\theta)\right)^2}.
\end{eqnarray}
where:
\begin{eqnarray}
    \Delta & = & 1 + \sum_{n=2}^{m} J_n \frac{\partial P_n(\sin(\theta)\sin(i))}{\partial (\sin(\theta)\sin(i))} \left(\sqrt{A}\right)^{n} \nonumber \\
	& \times & \displaystyle\frac{\cos^2(i)}{\sin(i)}\sin(\theta)(1 + e_x\cos(\theta) + e_y\sin(\theta))^{n-1}.
\end{eqnarray}



\section{Perturbation method}

The perturbation method used in this work follows the approach from Arnas~\cite{frozenj2}. Particularly, we know that frozen orbits appear either close to the critical inclination, or at small eccentricities with a magnitude of the order of $J_2$. In this work we first focus on the low eccentric frozen orbits and thus, it is possible to normalize the two components of the eccentricity vector in terms of the small parameter $J_2$. This is done via the following variable transformation:
\begin{eqnarray}
    X = \displaystyle\frac{e_x}{J_2}, \nonumber \\
    Y = \displaystyle\frac{e_y}{J_2}.
\end{eqnarray}
Additionally, for an Earth-like planet, we know that the terms $J_n$, with $n > 2$ are on the order of magnitude of $J_2^2$. Therefore, these coefficients of the zonal harmonics are also normalized such that:
\begin{equation}
    \mathcal{J}_n = \displaystyle\frac{J_n}{J_2^2}.
\end{equation}

Once this is done, a power series expansion in the small parameter $J_2$ is performed for all the dependent variables in the problem \{$A$, $X$, $Y$, $i$, $\Omega$, $t$\}:
\begin{eqnarray}
    A & \approx & A_0 + A_1 J_2 + A_2 J_2^2, \nonumber \\
    X & \approx & X_{1} + X_{2} J_2, \nonumber \\
    Y & \approx & Y_{1} + Y_{2} J_2, \nonumber \\
    i & \approx & i_0 + i_1 J_2 + i_2 J_2^2, \nonumber \\
    \Omega & \approx & \Omega_0 + \Omega_1 J_2 + \Omega_2 J_2^2, \nonumber \\
    t & \approx & t_0 + t_{1} J_2 + t_{2} J_2^2.
\end{eqnarray}
as well as in the differential equation itself. Note that variables $X$ and $Y$ have only be expanded up to order 1 since this represents an order 2 in the original variables $e_x$ and $e_y$. These expansions lead to a system of equations where it is possible to identify the parts of the differential equation that have the same order of magnitude based on the power series expansion. In particular, for the zero order solution (the unperturbed system) the following differential equation is obtained:
\begin{eqnarray}
    \displaystyle\frac{dA_0}{d\theta} & = & 0; \nonumber \\
    \displaystyle\frac{di_0}{d\theta} & = & 0; \nonumber \\
    \displaystyle\frac{d\Omega_0}{d\theta} & = & 0; \nonumber \\
    \displaystyle\frac{d t_0}{d\theta} & = & \sqrt[4]{\frac{R^6}{A_0^3 \mu^2}}.
\end{eqnarray}

For first order, the resultant system is the following:
\begin{eqnarray}
    \displaystyle\frac{dA_1}{d\theta} & = & 6 A_0^2 \sin ^2(i_0) \sin (2 \theta); \nonumber \\
    \displaystyle\frac{dX_1}{d\theta} & = & \frac{3}{2} A_0 \sin (\theta) \left(3 \sin ^2(i_0) \sin ^2(\theta)-4 \sin ^2(i_0) \cos
   ^2(\theta)-1\right); \nonumber \\
    \displaystyle\frac{dY_1}{d\theta} & = & \frac{3}{2} A_0 \cos (\theta) \left(1-7 \sin ^2(i_0) \sin ^2(\theta)\right); \nonumber \\
    \displaystyle\frac{di_1}{d\theta} & = & -3 A_0 \sin (i_0) \cos (i_0) \sin (\theta) \cos (\theta); \nonumber \\
    \displaystyle\frac{d\Omega_1}{d\theta} & = & -3 A_0 \cos (i_0) \sin ^2(\theta); \nonumber \\
    \displaystyle\frac{d t_1}{d\theta} & = & -\sqrt[4]{\frac{R^6}{A_0^3 \mu^2}} \Big[\frac{3}{4}\frac{A_1}{A_0} + 3 A_0 \cos
   ^2(i_0) \sin ^2(\theta) + (2 X_1 \cos (\theta) + 2 Y_1 \sin
   (\theta)) \Big];
\end{eqnarray}
where it is important to note that, as expected, the differential equation only depends on the terms related with $J_2$. Note also that the initial condition for \{$A_1$, $i_1$, $\Omega_1$ and $t_1$ is zero as they represent the first order deviation from the unperturbed problem. In contrast, the initial conditions of the normalized components of the eccentricity vector are:
\begin{eqnarray}
    X_1 (t=0) = X_0 = \displaystyle\frac{e_{x}(t = 0)}{J_2}, & & 
    Y_1 (t=0) = Y_0 = \displaystyle\frac{e_{y}(t = 0)}{J_2},
\end{eqnarray}
respectively.

For second order, it is possible to separate the contributions of each specific zonal harmonic since there are no mixed terms, and the differential equation depends only on the zero and first order solutions (with the exemption of the time evolution). Particularly, in this work we will focus on the $J_2$, $J_3$, $J_4$, $J_5$, $J_6$, and $J_7$ terms of the zonal harmonics. Therefore, the second term of each dependent variable can be further decomposed into:
\begin{eqnarray}
    A_2 & = & A_2|_{J_2} + A_2|_{J_3} + A_2|_{J_4} + A_2|_{J_5} + A_2|_{J_6} + A_2|_{J_7}, \nonumber \\
    X_2 & = & X_2|_{J_2} + X_2|_{J_3} + X_2|_{J_4} + X_2|_{J_5} + X_2|_{J_6} + X_2|_{J_7}, \nonumber \\
    Y_2 & = & Y_2|_{J_2} + Y_2|_{J_3} + Y_2|_{J_4} + Y_2|_{J_5} + Y_2|_{J_6} + Y_2|_{J_7}, \nonumber \\
    i_2 & = & i_2|_{J_2} + i_2|_{J_3} + i_2|_{J_4} + i_2|_{J_5} + i_2|_{J_6} + i_2|_{J_7}, \nonumber \\
    \Omega_2 & = & \Omega_2|_{J_2} + \Omega_2|_{J_3} + \Omega_2|_{J_4} + \Omega_2|_{J_5} + \Omega_2|_{J_6} + \Omega_2|_{J_7},
\end{eqnarray}
and identified with their correspondent part in the differential equation. Particularly, for $J_2$:
\begin{eqnarray}
    \displaystyle\frac{dA_2|_{J_2}}{d\theta} & = & 12 A_0 \sin (i_0) \sin (\theta) \cos (\theta) \Big(-3 A_0^2 \sin (i_0) \cos ^2(i_0) \sin
   ^2(\theta) \nonumber \\
   & + & \sin (i_0) (A_0 X_1 \cos (\theta)+A_0 Y_1 \sin (\theta)+2 A_1)+2 A_0 i_1
   \cos (i_0)\Big); \nonumber \\
    \displaystyle\frac{dX_2|_{J_2}}{d\theta} & = & -\frac{3}{2} \sin (\theta) \Big(-3 \sin ^2(\theta) \Big(A_0^2 \cos ^2(i_0)+A_0 i_1
   \sin (2 i_0)+A_1 \sin ^2(i_0)\Big) \nonumber \\
   & - & \frac{3}{4} A_0^2 \sin ^2(2 i_0)
   \sin ^2(2 \theta)+9 A_0^2 \sin ^2(i_0) \cos ^2(i_0) \sin ^4(\theta) \nonumber \\
   & + & 2 \sin (i_0) \cos
   ^2(\theta) (\sin (i_0) (3 A_0 Y_1 \sin (\theta)+2 A_1)+4 A_0 i_1 \cos (i_0)) \nonumber \\
    & + & 5
   A_0 X_1 \sin ^2(i_0) \cos ^3(\theta)+A_0 X_1 \cos (\theta) \Big(\sin ^2(i_0) \Big(3-7 \sin
   ^2(\theta)\Big)+2\Big) \nonumber \\
   & - & 6 A_0 Y_1 \sin ^2(i_0) \sin ^3(\theta)+A_0 Y_1 (\cos (2 i_0)+3)
   \sin (\theta)+A_1\Big); \nonumber \\
    \displaystyle\frac{dY_2|_{J_2}}{d\theta} & = & \frac{3}{2} \Big(\cos (\theta) \Big(3 A_0^2 \cos ^2(i_0) \sin ^2(\theta) \Big(7 \sin
   ^2(i_0) \sin ^2(\theta)-1\Big) \nonumber \\
   & - & 7 \sin ^2(i_0) \sin ^2(\theta) (2 A_0 Y_1 \sin
   (\theta)+A_1)-7 A_0 i_1 \sin (2 i_0) \sin ^2(\theta)+A_1\Big) \nonumber \\
   & + & A_0
   \Big(\sin (2 \theta) \Big(Y_1-3 X_1 \sin ^2(i_0) \sin (2 \theta)\Big)+2 X_1 \cos
   ^2(i_0) \sin ^2(\theta)\Big) \nonumber \\
   & - & 2 A_0 Y_1 \sin ^2(i_0) \sin (\theta) \cos ^3(\theta)+2 A_0 X_1 \cos
   ^2(\theta)\Big); \nonumber \\
    \displaystyle\frac{di_2|_{J_2}}{d\theta} & = & -3 \sin (\theta) \cos (\theta) \Big(-3 A_0^2 \sin (i_0) \cos ^3(i_0) \sin ^2(\theta)-A_0 i_1 \sin ^2(i_0) \nonumber \\
    & + & A_0
   i_1 \cos ^2(i_0) + \sin (i_0)
   \cos (i_0) (A_0 X_1 \cos (\theta)+A_0 Y_1 \sin (\theta)+A_1)\Big); \nonumber \\
    \displaystyle\frac{d\Omega_2|_{J_2}}{d\theta} & = & -3 \sin ^2(\theta) \Big(-3 A_0^2 \cos ^3(i_0) \sin ^2(\theta) \nonumber \\
    & + & \cos (i_0) (A_0 X_1 \cos (\theta)+A_0
   Y_1 \sin (\theta)+A_1)-A_0 i_1 \sin (i_0)\Big);
\end{eqnarray}
for $J_3$:
\begin{eqnarray}
    \displaystyle\frac{dA_2|_{J_3}}{d\theta} & = & 6 A_0^{5/2} \mathcal{J}_3 \sin (i_0) \cos (\theta) \Big(5 \sin ^2(i_0) \sin
   ^2(\theta)-1\Big); \nonumber \\
    \displaystyle\frac{dX_2|_{J_3}}{d\theta} & = & -A_0^{3/2} \mathcal{J}_3\sin (i_0) \Big(-10 \sin ^2(i_0) \sin ^4(\theta) \nonumber \\
    & + & 3 \cos ^2(\theta)
   \Big(5 \sin ^2(i_0) \sin ^2(\theta)-1\Big)+6 \sin ^2(\theta)\Big); \nonumber \\
    \displaystyle\frac{dY_2|_{J_3}}{d\theta} & = & A_0^{3/2} \mathcal{J}_3\sin (i_0) \sin (\theta) \cos (\theta) \Big(9-25 \sin ^2(i_0) \sin
   ^2(\theta)\Big); \nonumber \\
    \displaystyle\frac{di_2|_{J_3}}{d\theta} & = & \frac{3}{2} A_0^{3/2} \mathcal{J}_3 \cos (i_0) \cos (\theta) \Big(1-5 \sin ^2(i_0) \sin
   ^2(\theta)\Big); \nonumber \\
    \displaystyle\frac{d\Omega_2|_{J_3}}{d\theta} & = & \frac{3}{2} A_0^{3/2} \mathcal{J}_3 \sin (\theta) \Big(\cot (i_0)-5 \sin (i_0) \cos (i_0)
   \sin ^2(\theta)\Big);
\end{eqnarray}
for $J_4$:
\begin{eqnarray}
    \displaystyle\frac{dA_2|_{J_4}}{d\theta} & = & 5 A_0^3 \mathcal{J}_4 \sin ^2(i_0) \sin (2 \theta) \Big(7 \sin ^2(i_0) \sin
   ^2(\theta)-3\Big); \nonumber \\
    \displaystyle\frac{dX_2|_{J_4}}{d\theta} & = & \frac{1}{2} A_0^2 \mathcal{J}_4 \sin (\theta) \Big(35 \sin ^4(i_0) \sin ^4(\theta)-30 \sin
   ^2(i_0) \sin ^2(\theta) \nonumber \\
    & + & \cos ^2(\theta) \Big(30 \sin ^2(i_0)-70 \sin ^4(i_0) \sin
   ^2(\theta)\Big)+3\Big); \nonumber \\
    \displaystyle\frac{dY_2|_{J_4}}{d\theta} & = & -\frac{5}{8} A_0^2 \mathcal{J}_4 \cos (\theta) \Big(91 \sin ^4(i_0) \sin ^4(\theta)-54 \sin
   ^2(i_0) \sin ^2(\theta)+3\Big); \nonumber \\
    \displaystyle\frac{di_2|_{J_4}}{d\theta} & = & -\frac{5}{2} A_0^2 \mathcal{J}_4 \sin (i_0) \cos (i_0) \sin (\theta) \cos (\theta) \Big(7
   \sin ^2(i_0) \sin ^2(\theta)-3\Big); \nonumber \\
    \displaystyle\frac{d\Omega_2|_{J_3}}{d\theta} & = & -\frac{5}{2} A_0^2 \mathcal{J}_4 \cos (i_0) \sin ^2(\theta) \Big(7 \sin ^2(i_0) \sin
   ^2(\theta)-3\Big);
\end{eqnarray}
for $J_5$:
\begin{eqnarray}
    \displaystyle\frac{dA_2|_{J_5}}{d\theta} & = & \frac{15}{2} A_0^{7/2} \mathcal{J}_5 \sin (i_0) \cos (\theta) \Big(21 \sin ^4(i_0) \sin
   ^4(\theta)-14 \sin ^2(i_0) \sin ^2(\theta)+1\Big); \nonumber \\
    \displaystyle\frac{dX_2|_{J_5}}{d\theta} & = & -\frac{1}{4} A_0^{5/2} \mathcal{J}_5 \sin (i_0) \Big(15 \cos ^2(\theta) \Big(21 \sin
   ^4(i_0) \sin ^4(\theta)-14 \sin ^2(i_0) \sin ^2(\theta)+1\Big) \nonumber \\
    & - & 2 \sin ^2(\theta)
   \Big(63 \sin ^4(i_0) \sin ^4(\theta)-70 \sin ^2(i_0) \sin
   ^2(\theta)+15\Big)\Big); \nonumber \\
    \displaystyle\frac{dY_2|_{J_5}}{d\theta} & = & -3 A_0^{5/2} \mathcal{J}_5 \sin (i_0) \sin (\theta) \cos (\theta) \Big(42 \sin ^4(i_0) \sin
   ^4(\theta)-35 \sin ^2(i_0) \sin ^2(\theta)+5\Big); \nonumber \\
    \displaystyle\frac{di_2|_{J_5}}{d\theta} & = & -\frac{15}{8} A_0^{5/2} \mathcal{J}_5 \cos (i_0) \cos (\theta) \Big(21 \sin ^4(i_0) \sin
   ^4(\theta)-14 \sin ^2(i_0) \sin ^2(\theta)+1\Big); \nonumber \\
    \displaystyle\frac{d\Omega_2|_{J_3}}{d\theta} & = & -\frac{15}{8} A_0^{5/2} \mathcal{J}_5 \cot (i_0) \sin (\theta) \Big(21 \sin ^4(i_0) \sin
   ^4(\theta)-14 \sin ^2(i_0) \sin ^2(\theta)+1\Big);
\end{eqnarray}
for $J_6$:
\begin{eqnarray}
    \displaystyle\frac{dA_2|_{J_6}}{d\theta} & = & \frac{21}{2} A_0^4 \mathcal{J}_6 \sin ^2(i_0) \sin (\theta) \cos (\theta) \Big(33 \sin
   ^4(i_0) \sin ^4(\theta)-30 \sin ^2(i_0) \sin ^2(\theta)+5\Big); \nonumber \\
    \displaystyle\frac{dX_2|_{J_6}}{d\theta} & = & \frac{1}{4} A_0^3 \mathcal{J}_6 \sin (\theta) \Big(231 \sin ^6(i_0) \sin ^6(\theta)-315
   \sin ^4(i_0) \sin ^4(\theta)+105 \sin ^2(i_0) \sin ^2(\theta) \nonumber \\
    & - & 21 \sin ^2(i_0) \cos
   ^2(\theta) \Big(33 \sin ^4(i_0) \sin ^4(\theta)-30 \sin ^2(i_0) \sin
   ^2(\theta)+5\Big)-5\Big); \nonumber \\
    \displaystyle\frac{dY_2|_{J_6}}{d\theta} & = & -\frac{7}{16} A_0^3 \mathcal{J}_6 \cos (\theta) \Big(627 \sin ^6(i_0) \sin ^6(\theta)-675
   \sin ^4(i_0) \sin ^4(\theta) \nonumber \\
    & + & 165 \sin ^2(i_0) \sin ^2(\theta)-5\Big); \nonumber \\
    \displaystyle\frac{di_2|_{J_6}}{d\theta} & = & -\frac{21}{8} A_0^3 \mathcal{J}_6 \sin (i_0) \cos (i_0) \sin (\theta) \cos (\theta) \Big(33
   \sin ^4(i_0) \sin ^4(\theta) \nonumber \\
    & - & 30 \sin ^2(i_0) \sin ^2(\theta)+5\Big); \nonumber \\
    \displaystyle\frac{d\Omega_2|_{J_3}}{d\theta} & = & -\frac{21}{8} A_0^3 \mathcal{J}_6 \cos (i_0) \sin ^2(\theta) \Big(33 \sin ^4(i_0) \sin
   ^4(\theta)-30 \sin ^2(i_0) \sin ^2(\theta)+5\Big);
\end{eqnarray}
and for $J_7$:
\begin{eqnarray}
    \displaystyle\frac{dA_2|_{J_7}}{d\theta} & = & \frac{7}{4} A_0^{9/2} \mathcal{J}_7 \sin (i_0) \cos (\theta) \Big(429 \sin ^6(i_0) \sin
   ^6(\theta)-495 \sin ^4(i_0) \sin ^4(\theta) \nonumber \\
    & + & 135 \sin ^2(i_0) \sin ^2(\theta)-5\Big); \nonumber \\
    \displaystyle\frac{dX_2|_{J_7}}{d\theta} & = & -\frac{1}{8} A_0^{7/2} \mathcal{J}_7 \sin (i_0) \Big(-858 \sin ^6(i_0) \sin
   ^8(\theta)+1386 \sin ^4(i_0) \sin ^6(\theta) \nonumber \\
    & - & 630 \sin ^2(i_0) \sin ^4(\theta)+7 \cos ^2(\theta)
   \Big(429 \sin ^6(i_0) \sin ^6(\theta) \nonumber \\
    & - & 495 \sin ^4(i_0) \sin ^4(\theta)+135 \sin
   ^2(i_0) \sin ^2(\theta)-5\Big)+70 \sin ^2(\theta)\Big); \nonumber \\
    \displaystyle\frac{dY_2|_{J_7}}{d\theta} & = & \frac{1}{8} A_0^3 \mathcal{J}_7 \sin (i_0) \sin (\theta) \cos (\theta) \Big(-4719 \sin
   ^6(i_0) \sin ^6(\theta) \nonumber \\
    & + & 6237 \sin ^4(i_0) \sin ^4(\theta)-2205 \sin ^2(i_0) \sin
   ^2(\theta)+175\Big); \nonumber \\
    \displaystyle\frac{di_2|_{J_7}}{d\theta} & = & -\frac{7}{16} A_0^{7/2} \mathcal{J}_7 \cos (i_0) \cos (\theta) \Big(429 \sin ^6(i_0)
   \sin ^6(\theta) \nonumber \\
    & - & 495 \sin ^4(i_0) \sin ^4(\theta)+135 \sin ^2(i_0) \sin ^2(\theta)-5\Big); \nonumber \\
    \displaystyle\frac{d\Omega_2|_{J_7}}{d\theta} & = & -\frac{7}{16} A_0^{7/2} \mathcal{J}_7 \cot (i_0) \sin (\theta) \Big(429 \sin ^6(i_0)
   \sin ^6(\theta) \nonumber \\
    & - & 495 \sin ^4(i_0) \sin ^4(\theta)+135 \sin ^2(i_0) \sin ^2(\theta)-5\Big).
\end{eqnarray}
The time evolution, unfortunately, has to be treated separately as it depends on the second order solution of the orbital element $A$. More specifically, the second order differential equation for the time evolution is:
\begin{eqnarray}
    \displaystyle\frac{dt_2}{d\theta} & = & \sqrt[4]{\frac{R^6}{A_0^3 \mu^2}} \Big(240 A_0^{7/2} \mathcal{J}_3 \sin (i_0)
   \cos ^2(i_0) \sin ^3(\theta)-48 A_0^{7/2} \mathcal{J}_3 \cos (i_0) \cot (i_0) \sin
   (\theta) \nonumber \\
    & - & 840 A_0^{9/2} \mathcal{J}_5 \sin (i_0) \cos ^2(i_0) \sin ^3(\theta)+1260 A_0^{9/2}
   \mathcal{J}_5 \sin ^3(i_0) \cos ^2(i_0) \sin ^5(\theta) \nonumber \\
    & + & 60 A_0^{9/2} \mathcal{J}_5 \cos
   (i_0) \cot (i_0) \sin (\theta)+1890 A_0^{11/2} \mathcal{J}_7 \sin (i_0) \cos ^2(i_0) \sin
   ^3(\theta) \nonumber \\
    & + & 6006 A_0^{11/2} \mathcal{J}_7 \sin ^5(i_0) \cos ^2(i_0) \sin ^7(\theta)-6930
   A_0^{11/2} \mathcal{J}_7 \sin ^3(i_0) \cos ^2(i_0) \sin ^5(\theta) \nonumber \\
    & - & 70 A_0^{11/2}
   \mathcal{J}_7 \cos (i_0) \cot (i_0) \sin (\theta)+420 A_0^5 \mathcal{J}_6 \cos ^2(i_0) \sin
   ^2(\theta) \nonumber \\
    & + & 2772 A_0^5 \mathcal{J}_6 \sin ^4(i_0) \cos ^2(i_0) \sin ^6(\theta)-2520 A_0^5
   \mathcal{J}_6 \sin ^2(i_0) \cos ^2(i_0) \sin ^4(\theta) \nonumber \\
    & - & 240 A_0^4 \mathcal{J}_4 \cos ^2(i_0)
   \sin ^2(\theta)+560 A_0^4 \mathcal{J}_4 \sin ^2(i_0) \cos ^2(i_0) \sin ^4(\theta) \nonumber \\
    & - & 288 A_0^4
   \cos ^4(i_0) \sin ^4(\theta)-192 A_0^3 i_1 \sin (i_0) \cos (i_0) \sin ^2(\theta) \nonumber \\
    & - & 96
   A_0^3 X_1 \cos ^2(i_0) \sin ^2(\theta) \cos (\theta)-96 A_0^3 Y_1 \cos ^2(i_0) \sin ^3(\theta) \nonumber \\
    & + & 24
   A_0^2 A_1 \cos ^2(i_0) \sin ^2(\theta)-96 A_0^2 X_1^2 \cos ^2(\theta)-192 A_0^2 X_1 Y_1
   \sin (\theta) \cos (\theta) \nonumber \\
    & + & 64 A_0^2 X_2 \cos (\theta)-96 A_0^2 Y_1^2 \sin ^2(\theta)+64
   A_0^2 Y_2 \sin (\theta) \nonumber \\
    & - & 48 A_0 A_1 X_1 \cos (\theta)-48 A_0 A_1 Y_1 \sin
   (\theta)+24 A_0 A_2-21 A_1^2\Big).
\end{eqnarray}

In the following sections, the solutions to each of these individual systems are provided, as well as the related transformation from osculating to mean elements, and the secular variation of each orbital element.


\section{Approximate analytic solution}

One of the advantages of applying the proposed perturbation methodology is that the approximate analytical solution of the problem can be obtained by direct integration of the differential equations presented in the previous section. In particular, and for a general dependent variable $\phi$ of order $n$, the osculating evolution of it is given by:
\begin{equation}
    \phi_n = \int_{\theta_0}^{\theta} \displaystyle\frac{d \phi_n}{d\theta},
\end{equation}
being the actual osculating value of the variable, the addition of each term of the power series:
\begin{equation}
    \phi = \sum_{n=0}^{m} \phi_n J_2^n.
\end{equation}

\subsection{Zero order and first order solutions}

The zero and first order solutions of the zonal harmonics problem correspond to the zero and first order solutions of the $J_2$ problem. This means that these solutions are exactly the same ones already provided by Arnas~\cite{frozenj2}. Therefore, they are not repeated in this manuscript, so the author invites the reader to access Ref.~\cite{frozenj2} to obtain this information.

\subsection{Second order solution}

One of the advantages of separating the second order solution into the different contributions of the zonal harmonics terms is that its solution can be expressed as a combination of each individual component. As a result of that, the contribution, in second order, from the $J_2$ term of the potential, is the same as the one obtained in Ref.~\cite{frozenj2}. Thus, it is only necessary to focus on the contributions of the rest of zonal harmonics terms in this manuscript. Note that the expressions of the time evolution solution are not included in this document due to their length, however, they can be found in:
\href{https://engineering.purdue.edu/ART/research/research-code}{https://engineering.purdue.edu/ART/research/research-code}
with the rest of the code associated with this work.

The second order effects of $J_3$ are:
\begin{eqnarray}
    A_2|_{J_3} & = & 2 A_0^{5/2} \mathcal{J}_3 \sin (i_0) \big(5 \sin ^2(i_0) \big(\sin
   ^3(\theta)-\sin ^3(\theta_0)\big)+3 (\sin (\theta_0)-\sin
   (\theta))\big); \nonumber \\
    X_2|_{J_3} & = & -\frac{1}{32} A_0^{3/2} \mathcal{J}_3 \sin (i_0) \big(5 \sin ^2(i_0) (12
   \theta_0 - 16 \sin (2 \theta_0)+5 \sin (4 \theta_0)-12 \theta \nonumber \\
   & + & 16
   \sin (2 \theta)-5 \sin (4 \theta))-48 (\theta_0-3 \sin
   (\theta_0) \cos (\theta_0)-\theta+3 \sin (\theta) \cos
   (\theta))\big);\nonumber \\
    Y_2|_{J_3} & = & \frac{1}{4} A_0^{3/2} \mathcal{J}_3 \sin (i_0) \big(25 \sin ^2(i_0) \big(\sin
   ^4(\theta_0)-\sin ^4(\theta)\big)+9 \cos (2 \theta_0)-9 \cos (2
   \theta)\big); \nonumber \\
    i_2|_{J_3} & = & \frac{1}{2} A_0^{3/2} \mathcal{J}_3 \cos (i_0) \big(5 \sin ^2(i_0) \big(\sin
   ^3(\theta_0)-\sin ^3(\theta)\big)+3 (\sin (\theta)-\sin
   (\theta_0))\big);\nonumber \\
    \Omega_2|_{J_3} & = & \frac{1}{16} A_0^{3/2} \mathcal{J}_3 (5 \sin (2 i_0) (-9 \cos (\theta_0)+\cos (3
   \theta_0)+9 \cos (\theta)-\cos (3 \theta)) \nonumber \\
   & + & 24 \cot (i_0) (\cos
   (\theta_0)-\cos (\theta)));
\end{eqnarray}
for $J_4$:
\begin{eqnarray}
    A_2|_{J_4} & = & -\frac{5}{2} A_0^3 \mathcal{J}_4 \sin ^2(i_0) \big(7 \sin ^2(i_0) \big(\sin
   ^4(\theta_0)-\sin ^4(\theta)\big)+3 \cos (2 \theta_0)-3 \cos (2
   \theta)\big); \nonumber \\
    X_2|_{J_4} & = & \frac{1}{96} A_0^2 \mathcal{J}_4 \big(7 \sin ^4(i_0) (90 \cos (\theta_0)-35 \cos
   (3 \theta_0)+9 \cos (5 \theta_0)-90 \cos (\theta) \nonumber \\
   & + & 35 \cos (3
   \theta)-9 \cos (5 \theta))+240 \sin ^2(i_0) (-3 \cos
   (\theta_0)+\cos (3 \theta_0)+3 \cos (\theta) \nonumber \\
   & - & \cos (3
   \theta))+144 (\cos (\theta_0)-\cos (\theta))\big);\nonumber \\
    Y_2|_{J_4} & = & \frac{1}{8} A_0^2 \mathcal{J}_4 \big(91 \sin ^4(i_0) \big(\sin
   ^5(\theta_0)-\sin ^5(\theta)\big)+90 \sin ^2(i_0) \big(\sin
   ^3(\theta)-\sin ^3(\theta_0)\big) \nonumber \\
   & + & 15 (\sin (\theta_0)-\sin
   (\theta))\big); \nonumber \\
    i_2|_{J_4} & = & \frac{5}{8} A_0^2 \mathcal{J}_4 \sin (i_0) \cos (i_0) \big(7 \sin ^2(i_0) \big(\sin
   ^4(\theta_0)-\sin ^4(\theta)\big)+3 \cos (2 \theta_0)-3 \cos (2
   \theta)\big);\nonumber \\
    \Omega_2|_{J_4} & = & \frac{5}{64} A_0^2 \mathcal{J}_4 \cos (i_0) \big(7 \sin ^2(i_0) (12 \theta_0-8
   \sin (2 \theta_0)+\sin (4 \theta_0)-12 \theta+8 \sin (2
   \theta) \nonumber \\
   & - & \sin (4 \theta))+48 (-\theta_0+\sin (\theta_0) \cos
   (\theta_0)+\theta-\sin (\theta) \cos (\theta))\big);
\end{eqnarray}
for $J_5$:
\begin{eqnarray}
    A_2|_{J_5} & = & \frac{1}{2} A_0^{7/2} \mathcal{J}_5 \sin (i_0) \big(63 \sin ^4(i_0) \big(\sin
   ^5(\theta)-\sin ^5(\theta_0)\big) \nonumber \\
   & + & 70 \sin ^2(i_0) \big(\sin
   ^3(\theta_0)-\sin ^3(\theta)\big)+15 (\sin (\theta)-\sin
   (\theta_0))\big); \nonumber \\
    X_2|_{J_5} & = & -\frac{1}{256} A_0^{5/2} \mathcal{J}_5 \sin (i_0) \big(21 \sin ^4(i_0) (60
   \theta_0-75 \sin (2 \theta_0)+33 \sin (4 \theta_0) \nonumber \\
   & - & 7 \sin (6
   \theta_0)-60 \theta+75 \sin (2 \theta)-33 \sin (4 \theta)+7
   \sin (6 \theta))-140 \sin ^2(i_0) (12 \theta_0 \nonumber \\
   & - & 16 \sin (2
   \theta_0)+5 \sin (4 \theta_0)-12 \theta+16 \sin (2 \theta)-5
   \sin (4 \theta)) \nonumber \\
   & + & 480 (\theta_0-3 \sin (\theta_0) \cos
   (\theta_0)-\theta+3 \sin (\theta) \cos (\theta))\big);\nonumber \\
    Y_2|_{J_5} & = & \frac{3}{4} A_0^{5/2} \mathcal{J}_5 \sin (i_0) \big(28 \sin ^4(i_0) \big(\sin
   ^6(\theta_0)-\sin ^6(\theta)\big) \nonumber \\
   & + & 35 \sin ^2(i_0) \big(\sin
   ^4(\theta)-\sin ^4(\theta_0)\big) - 5 \cos (2 \theta_0)+5 \cos (2
   \theta)\big); \nonumber \\
    i_2|_{J_5} & = & \frac{1}{8} A_0^{5/2} \mathcal{J}_5 \cos (i_0) \big(63 \sin ^4(i_0) \big(\sin
   ^5(\theta_0)-\sin ^5(\theta)\big) \nonumber \\
   & + & 70 \sin ^2(i_0) \big(\sin
   ^3(\theta)-\sin ^3(\theta_0)\big)+15 (\sin (\theta_0)-\sin
   (\theta))\big);\nonumber \\
    \Omega_2|_{J_5} & = & \frac{1}{128} A_0^{5/2} \mathcal{J}_5 \big(-21 \sin ^3(i_0) \cos (i_0) (150 \cos
   (\theta_0)-25 \cos (3 \theta_0)+3 \cos (5 \theta_0) \nonumber \\
   & - & 150 \cos
   (\theta)+25 \cos (3 \theta)-3 \cos (5 \theta))+280 \sin (i_0)
   \cos (i_0) (9 \cos (\theta_0) \nonumber \\
   & - & \cos (3 \theta_0)-9 \cos (\theta)+\cos
   (3 \theta))-240 \cot (i_0) (\cos (\theta_0)-\cos (\theta))\big);
\end{eqnarray}
for $J_6$:
\begin{eqnarray}
    A_2|_{J_6} & = & \frac{21}{8} A_0^4 \mathcal{J}_6 \sin ^2(i_0) \big(22 \sin ^4(i_0) \big(\sin
   ^6(\theta)-\sin ^6(\theta_0)\big) \nonumber \\
   & + & 30 \sin ^2(i_0) \big(\sin
   ^4(\theta_0)-\sin ^4(\theta)\big)+5 (\cos (2 \theta_0)-\cos (2
   \theta))\big); \nonumber \\
    X_2|_{J_6} & = & \frac{1}{1280}A_0^3 \mathcal{J}_6 \big(-264 \sin ^6(i_0) \big((-108 \cos (2
   \theta_0)+15 \cos (4 \theta_0)+157) \cos ^3(\theta_0) \nonumber \\
   & + & \cos
   ^3(\theta) (108 \cos (2 \theta)-15 \cos (4
   \theta)-157)\big) \nonumber \\
   & + & 6720 \sin ^4(i_0) \big((7-3 \cos (2 \theta_0))
   \cos ^3(\theta_0) + \cos ^3(\theta) (3 \cos (2
   \theta)-7)\big) \nonumber \\
   & - & 11200 \sin ^2(i_0) \big(\cos ^3(\theta_0)-\cos
   ^3(\theta)\big) + 33 \sin ^6(i_0) (1225 \cos (\theta_0) \nonumber \\
   & - & 245 \cos (3
   \theta_0)+49 \cos (5 \theta_0)-5 \cos (7 \theta_0)-1225 \cos
   (\theta) + 245 \cos (3 \theta) \nonumber \\
   & - & 49 \cos (5 \theta)+5 \cos (7
   \theta))-420 \sin ^4(i_0) (150 \cos (\theta_0)-25 \cos (3
   \theta_0) \nonumber \\
   & + & 3 \cos (5 \theta_0)-150 \cos (\theta)+25 \cos (3
   \theta)-3 \cos (5 \theta))+2800 \sin ^2(i_0) (9 \cos
   (\theta_0) \nonumber \\
   & - & \cos (3 \theta_0)-9 \cos (\theta)+\cos (3
   \theta))-1600 (\cos (\theta_0)-\cos (\theta))\big);\nonumber \\
    Y_2|_{J_6} & = & \frac{1}{16} A_0^3 \mathcal{J}_6 \big(627 \sin ^6(i_0) \big(\sin
   ^7(\theta_0)-\sin ^7(\theta)\big)+945 \sin ^4(i_0) \big(\sin
   ^5(\theta)-\sin ^5(\theta_0)\big) \nonumber \\
   & + & 385 \sin ^2(i_0) \big(\sin
   ^3(\theta_0)-\sin ^3(\theta)\big)+35 (\sin (\theta)-\sin
   (\theta_0))\big); \nonumber \\
    i_2|_{J_6} & = & \frac{21}{64} A_0^3 \mathcal{J}_6 \big(44 \sin ^5(i_0) \cos (i_0) \big(\sin
   ^6(\theta_0)-\sin ^6(\theta)\big) \nonumber \\
   & + & 60 \sin ^3(i_0) \cos (i_0)
   \big(\sin ^4(\theta)-\sin ^4(\theta_0)\big)-5 \sin (2 i_0) (\cos
   (2 \theta_0)-\cos (2 \theta))\big);\nonumber \\
    \Omega_2|_{J_6} & = & \frac{21}{512} A_0^3 \mathcal{J}_6 \cos (i_0) \big(11 \sin ^4(i_0) (60 \theta_0-45
   \sin (2 \theta_0)+9 \sin (4 \theta_0)-\sin (6 \theta_0) \nonumber \\
   & - & 60
   \theta+45 \sin (2 \theta)-9 \sin (4 \theta)+\sin (6
   \theta))+60 \sin ^2(i_0) (-12 \theta_0+8 \sin (2 \theta_0) \nonumber \\
   & - & \sin (4
   \theta_0)+12 \theta-8 \sin (2 \theta)+\sin (4 \theta)) \nonumber \\
   & + & 160
   (\theta_0-\sin (\theta_0) \cos (\theta_0)-\theta+\sin (\theta)
   \cos (\theta))\big);
\end{eqnarray}
and for $J_7$:
\begin{eqnarray}
    A_2|_{J_7} & = & \frac{1}{4} A_0^{9/2} \mathcal{J}_7 \sin (i_0) \big(429 \sin ^6(i_0) \big(\sin
   ^7(\theta)-\sin ^7(\theta_0)\big) \nonumber \\
   & + & 693 \sin ^4(i_0) \big(\sin
   ^5(\theta_0)-\sin ^5(\theta)\big) \nonumber \\
   & + & 315 \sin ^2(i_0) \big(\sin
   ^3(\theta)-\sin ^3(\theta_0)\big)+35 (\sin (\theta_0)-\sin
   (\theta))\big); \nonumber \\
    X_2|_{J_7} & = & -\frac{1}{8192}A_0^{7/2} \mathcal{J}_7 \sin (i_0) \big(143 \sin ^6(i_0) (840 \theta_0-1008
   \sin (2 \theta_0)+504 \sin (4 \theta_0) \nonumber \\
   & - & 176 \sin (6 \theta_0)+27
   \sin (8 \theta_0)-840 \theta+1008 \sin (2 \theta)-504 \sin (4
   \theta) \nonumber \\
   & + & 176 \sin (6 \theta) - 27 \sin (8 \theta))+3696 \sin
   ^4(i_0) (-60 \theta_0+75 \sin (2 \theta_0) \nonumber \\
   & - & 33 \sin (4 \theta_0)+7 \sin
   (6 \theta_0)+60 \theta - 75 \sin (2 \theta)+33 \sin (4
   \theta)-7 \sin (6 \theta)) \nonumber \\
   & + & 10080 \sin ^2(i_0) (12 \theta_0-16 \sin
   (2 \theta_0)+5 \sin (4 \theta_0) - 12 \theta+16 \sin (2 \theta) \nonumber \\
   & - & 5
   \sin (4 \theta))+8960 (-2 \theta_0+3 \sin (2 \theta_0)+2
   \theta-3 \sin (2 \theta))\big);\nonumber \\
    Y_2|_{J_7} & = & \frac{1}{64} A_0^{7/2} \mathcal{J}_7 \sin (i_0) \big(4719 \sin ^6(i_0) \big(\sin
   ^8(\theta_0)-\sin ^8(\theta)\big) \nonumber \\
   & + & 8316 \sin ^4(i_0) \big(\sin
   ^6(\theta)-\sin ^6(\theta_0)\big) \nonumber \\
   & + & 4410 \sin ^2(i_0) \big(\sin
   ^4(\theta_0)-\sin ^4(\theta)\big)+350 (\cos (2 \theta_0)-\cos (2
   \theta))\big); \nonumber \\
    i_2|_{J_7} & = & \frac{1}{16} A_0^{7/2} \mathcal{J}_7 \cos (i_0) \big(429 \sin ^6(i_0) \big(\sin
   ^7(\theta_0)-\sin ^7(\theta)\big) \nonumber \\
   & + & 693 \sin ^4(i_0) \big(\sin
   ^5(\theta)-\sin ^5(\theta_0)\big) \nonumber \\
   & + & 315 \sin ^2(i_0) \big(\sin
   ^3(\theta_0)-\sin ^3(\theta)\big)+35 (\sin (\theta)-\sin
   (\theta_0))\big);\nonumber \\
    \Omega_2|_{J_7} & = & \frac{1}{5120}A_0^{7/2} \mathcal{J}_7 \big(429 \sin ^5(i_0) \cos (i_0) (-1225 \cos
   (\theta_0)+245 \cos (3 \theta_0) \nonumber \\
   & - & 49 \cos (5 \theta_0)+5 \cos (7
   \theta_0)+1225 \cos (\theta)-245 \cos (3 \theta)+49 \cos (5
   \theta)\nonumber \\
   & - & 5 \cos (7 \theta))+4620 \sin ^3(i_0) \cos (i_0) (150 \cos
   (\theta_0)-25 \cos (3 \theta_0)+3 \cos (5 \theta_0) \nonumber \\
   & - & 150 \cos
   (\theta)+25 \cos (3 \theta)-3 \cos (5 \theta))+12600 \sin (2
   i_0) (-9 \cos (\theta_0) \nonumber \\
   & + & \cos (3 \theta_0)+9 \cos (\theta)-\cos (3
   \theta))+11200 \cot (i_0) (\cos (\theta_0)-\cos
   (\theta))\big);
\end{eqnarray}


\section{Transformation from osculating to mean elements} \label{sec:mean}

Based on the analytical approximate solution obtained in the previous section, it is possible to provide the value of the mean elements for any initial condition. This can be done by performing the integral of the osculating value for a complete nodal period, that is, from ($\theta_0-\pi$) to ($\theta_0+\pi$). Note that the initial point has been selected to be in the center of the domain of integration to account for the expected long-term secular effects in the orbital elements. In other words, the mean value of order $n$ of a general orbital element $\phi$ is:
\begin{equation}
    \overline{\phi_n} = \displaystyle\frac{1}{2\pi} \int_{\theta_0 - \pi}^{\theta_0 + \pi} \phi_n,
\end{equation}
being the mean value of the variable obtained by the addition of the power series:
\begin{equation}
    \overline{\phi} = \phi_0 + \sum_{n=1}^{m} \overline{\phi_n} J_2^n.
\end{equation}

As in the case of the analytical approximate solution, zero and first order are equal to the $J_2$ solution (the other zonal harmonics do not have contributions at these orders of solution). Additionally, and as done in the previous section, the contributions of the second order solution can be separated into the different terms of the zonal harmonics. Particularly, for $J_3$:
\begin{eqnarray}
    \overline{A_2|_{J_3}} & = & 2 A^{5/2} \mathcal{J}_3 \sin (i) \sin (\theta_0) (3 - 5 \sin^2(i) \sin^2(\theta_0)); \nonumber \\
    \overline{X_2|_{J_3}} & = & -\frac{1}{32} A^{3/2} \mathcal{J}_3 \sin (i) \big(25 \sin ^2(i) \sin (4
   \theta_0)+8 (5 \cos (2 i)+4) \sin (2 \theta_0)\big);\nonumber \\
    \overline{Y_2|_{J_3}} & = & \frac{1}{32} A^{3/2} \mathcal{J}_3 \sin (i) \big(25 \sin ^2(i) \big(8 \sin
   ^4(\theta_0)-3\big)+72 \cos (2 \theta_0)\big); \nonumber \\
    \overline{i_2|_{J_3}} & = & \frac{1}{2} A^{3/2} \mathcal{J}_3 \cos (i) \sin (\theta_0) \big(5 \sin ^2(i)
   \sin ^2(\theta_0)-3\big);\nonumber \\
    \overline{\Omega_2|_{J_3}} & = & \frac{1}{16} A^{3/2} \mathcal{J}_3 (5 \sin (2 i) \cos (3 \theta_0)+3 \cos
   (\theta_0) (8 \cot (i)-15 \sin (2 i)));
\end{eqnarray}
for $J_4$:
\begin{eqnarray}
    \overline{A_2|_{J_4}} & = & -\frac{5}{16} A^3 \mathcal{J}_4 \sin^2 (i) (24 \cos (2 \theta) + 7 \sin^2 (i) (-3 + 8 \sin^4(\theta_0))); \nonumber \\
    \overline{X_2|_{J_4}} & = & \frac{1}{96} A^2 \mathcal{J}_4 \big(7 \sin ^4(i) (90 \cos (\theta_0)-35 \cos
   (3 \theta_0)+9 \cos (5 \theta_0)) \nonumber \\
   & + & 240 \sin ^2(i) (\cos (3
   \theta_0)-3 \cos (\theta_0))+144 \cos (\theta_0)\big);\nonumber \\
    \overline{Y_2|_{J_4}} & = & \frac{1}{8} A^2 \mathcal{J}_4 \sin (\theta_0) \big(91 \sin ^4(i) \sin
   ^4(\theta_0)-90 \sin ^2(i) \sin ^2(\theta_0)+15\big); \nonumber \\
    \overline{i_2|_{J_4}} & = & \frac{5}{64} A^2 \mathcal{J}_4 \sin (i) \cos (i) \big(7 \sin ^2(i) \big(8
   \sin ^4(\theta_0)-3\big)+24 \cos (2 \theta_0)\big);\nonumber \\
    \overline{\Omega_2|_{J_4}} & = & \frac{5}{64} A^2 \mathcal{J}_4 \cos (i) \big(7 \sin ^2(i) (\sin (4
   \theta_0)-8 \sin (2 \theta_0))+48 \sin (\theta_0) \cos
   (\theta_0)\big);
\end{eqnarray}
for $J_5$:
\begin{eqnarray}
    \overline{A_2|_{J_5}} & = & \frac{1}{2} A^{7/2} \mathcal{J}_5 \sin (i) \sin (\theta_0) (-15 + 70 \sin^2 (i) \sin^2 (\theta_0) - 63 \sin^4 (i) \sin^4 (\theta_0)); \nonumber \\
    \overline{X_2|_{J_5}} & = & \frac{1}{256} A^{5/2} \mathcal{J}_5 \sin (i) \big(21 \sin ^4(i) (7 \sin (6
   \theta_0)-33 \sin (4 \theta_0)) \nonumber \\
   & + & 700 \sin ^2(i) \sin (4 \theta_0)+5
   \big(315 \sin ^4(i)-448 \sin ^2(i)+144\big) \sin (2
   \theta_0)\big);\nonumber \\
    \overline{Y_2|_{J_5}} & = & \frac{3}{32} A^{5/2} \mathcal{J}_5 \sin (i) \big(7 \sin ^2(i) \big(-40 \cos (2 \theta_0) \nonumber \\
   & + & 2 \sin
   ^2(i) \big(16 \sin ^6(\theta_0)-5\big)-40 \sin
   ^4(\theta_0)+15\big)\big); \nonumber \\
    \overline{i_2|_{J_5}} & = & \frac{1}{8} A^{5/2} \mathcal{J}_5 \cos (i) \sin (\theta_0) \big(63 \sin
   ^4(i) \sin ^4(\theta_0)-70 \sin ^2(i) \sin ^2(\theta_0)+15\big);\nonumber \\
    \overline{\Omega_2|_{J_5}} & = & \frac{1}{256 }A^{5/2} \mathcal{J}_5 \big(-42  \sin ^3(i) \cos (i) (150 \cos
   (\theta_0)-25 \cos (3 \theta_0)+3 \cos (5 \theta_0)) \nonumber \\
   & - & 280  \sin
   (2 i) (\cos (3 \theta_0)-9 \cos (\theta_0))-480  \cot (i) \cos
   (\theta_0)\big);
\end{eqnarray}
for $J_6$:
\begin{eqnarray}
    \overline{A_2|_{J_6}} & = & \frac{21}{64} A^4 \mathcal{J}_6 \sin ^2(i)  (40 \cos (2 \theta_0) + 
   \sin ^2(i)  (30 (-3 + 8 \sin ^4(\theta_0)) \nonumber \\
   & + & \sin ^2(i)  (55 - 176 \sin ^6(\theta_0)))); \nonumber \\
    \overline{X_2|_{J_6}} & = & \frac{1}{10240}A^3 \mathcal{J}_6 \big(8 \sin ^2(i) \big(-35 (78 \cos (2 i)+33 \cos (4
   i)+49) \cos (3 \theta_0) \nonumber \\
   & - & 660 \sin ^4(i) \cos (7 \theta_0)-84 \sin
   ^2(i) (22 \cos (2 i)+23) \cos (5 \theta_0)\big) \nonumber \\
   & - & 25 (105 \cos (2
   i)+126 \cos (4 i)+231 \cos (6 i)+50) \cos (\theta_0)\big);\nonumber \\
    \overline{Y_2|_{J_6}} & = & \frac{1}{16} A^3 \mathcal{J}_6 \sin (\theta_0) \big(627 \sin ^6(i) \sin
   ^6(\theta_0)-945 \sin ^4(i) \sin ^4(\theta_0) \nonumber \\
   & + & 385 \sin ^2(i) \sin
   ^2(\theta_0)-35\big); \nonumber \\
    \overline{i_2|_{J_6}} & = & \frac{21}{256} A^3 \mathcal{J}_6 \big(\sin ^3(i) \cos (i) \big(11 \sin ^2(i)
   \big(16 \sin ^6(\theta_0)-5\big)-240 \sin
   ^4(\theta_0)+90\big) \nonumber \\
   & - & 20 \sin (2 i) \cos (2 \theta_0)\big);\nonumber \\
    \overline{\Omega_2|_{J_6}} & = & -\frac{21}{512} A^3 \mathcal{J}_6 \cos (i) \big(11 \sin ^4(i) (\sin (6
   \theta_0)-9 \sin (4 \theta_0))+60 \sin ^2(i) \sin (4 \theta_0) \nonumber \\
   & + & 5
   \big(99 \sin ^4(i)-96 \sin ^2(i)+16\big) \sin (2 \theta_0)\big);
\end{eqnarray}
and for $J_7$:
\begin{eqnarray}
    \overline{A_2|_{J_7}} & = & -\frac{1}{4} A^{9/2} \mathcal{J}_7 \sin (i)  \sin (\theta_0) (-35 + 315 \sin ^2(i)  \sin^2 (\theta_0)  - 693 \sin ^4(i)  \sin^4 (\theta_0) \nonumber\\
    & + &   429 \sin ^6(i)  \sin^6 (\theta_0)); \nonumber \\
    \overline{X_2|_{J_7}} & = & \frac{1}{8192} A^{7/2} \mathcal{J}_7 \sin (i) \big(-143 \sin ^6(i) (504 \sin (4
   \theta_0)-176 \sin (6 \theta_0)+27 \sin (8 \theta_0)) \nonumber \\
   & + & 3696 \sin
   ^4(i) (33 \sin (4 \theta_0)-7 \sin (6 \theta_0))-50400 \sin ^2(i)
   \sin (4 \theta_0) \nonumber \\
   & + & 336 \big(429 \sin ^6(i)-825 \sin ^4(i)+480 \sin
   ^2(i)-80\big) \sin (2 \theta_0)\big);\nonumber \\
    \overline{Y_2|_{J_7}} & = & \frac{1}{64}A^{7/2} \mathcal{J}_7 \sin (i) \big (44800 \cos (2 \theta_0) + 
  3 \sin^2 (i) (23520 (-3 + 8 \sin^4 (\theta_0)) \nonumber \\
   & - &  22176 \sin^2 (i) (-5 + 16 \sin^6 (\theta_0)) + 
     1573 \sin^4 (i) (-35 + 128 \sin^8 (\theta_0)))\big); \nonumber \\
    \overline{i_2|_{J_7}} & = & \frac{1}{16} A^{7/2} \mathcal{J}_7 \cos (i) \sin (\theta_0) \big(429 \sin
   ^6(i) \sin ^6(\theta_0)-693 \sin ^4(i) \sin ^4(\theta_0) \nonumber \\
   & + & 315 \sin
   ^2(i) \sin ^2(\theta_0)-35\big);\nonumber \\
    \overline{\Omega_2|_{J_7}} & = & \frac{1}{5120}A^{7/2} \mathcal{J}_7 \big(429 \sin ^5(i) \cos (i) (-1225 \cos
   (\theta_0)+245 \cos (3 \theta_0)-49 \cos (5 \theta_0) \nonumber \\
   & + & 5 \cos (7
   \theta_0))+4620 \sin ^3(i) \cos (i) (150 \cos (\theta_0)-25 \cos (3
   \theta_0)+3 \cos (5 \theta_0)) \nonumber \\
   & + & 12600 \sin (2 i) (\cos (3 \theta_0)-9
   \cos (\theta_0))+11200 \cot (i) \cos (\theta_0)\big).
\end{eqnarray}


\section{Secular variation} \label{sec:secular}

The study of the secular variation of the orbital elements has three main goals. First, it provides the long-term rate of change of the variables involved in the problem. Second, it allows to obtain the perturbed period of the orbit for any initial condition. And third, it provides the conditions that allow to frozen an orbit. 

The secular variation in this work is considered as the variation that any of the variables has suffered after one complete orbital revolution. In other words, this is equivalent to imposing that $\theta = 2\pi$ in the approximate analytical solution obtained for the zonal harmonics problem. As in previous cases, it is possible to separate the effects of the different orders of the solution.

\subsection{First order secular variation}
The first order secular variation is only dependent on the effects of the $J_2$ term of the zonal harmonics, therefore, the result is the same as in Ref.~\cite{frozenj2}. Nevertheless, we include this result in here for completeness:
\begin{eqnarray}
\Delta\left.A_1\right|_{sec} & = & 0 \nonumber \\
\Delta\left.X_1\right|_{sec} & = & 0 \nonumber \\
\Delta\left.Y_1\right|_{sec} & = & 0 \nonumber \\
\Delta\left.i_1\right|_{sec} & = & 0 \nonumber \\
\Delta\left.\Omega_1\right|_{sec} & = & -3 A_0 \pi \cos(i_0)
\end{eqnarray}

\subsection{Second order secular variation}
The second order secular variation, on the other hand, depends on the effects of all the terms of the zonal harmonics. Particularly:
\begin{eqnarray}
\Delta\left.A_2\right|_{sec} & = & 0; \nonumber \\
\Delta\left.X_2\right|_{sec} & = & \displaystyle\frac{3}{128} A_0 \pi (-32 Y_0 (3 + 5 \cos(2 i_0)) + 
   3 A_0 (41 + 52 \cos(2 i_0) \nonumber \\
   & + & 35 \cos(4 i_0)) \sin(\theta_0) + 
   28 A_0 (3 + 5 \cos(2 i_0)) \sin^2(i_0) \sin(3 \theta_0))  \nonumber \\
   & - & \frac{3}{8} \pi  A_0^{3/2} \mathcal{J}_3 \sin (i_0) (5 \cos (2 i_0)+3)  \nonumber \\
   & + &  \frac{15}{512} \pi  A_0^{5/2} \mathcal{J}_5 (2 \sin (i_0)+7 (\sin (3 i_0)+3 \sin (5
   i_0)))  \nonumber \\
   & - &  \frac{35}{32768} \pi  A_0^{7/2} \mathcal{J}_7 (25 \sin (i_0)+81 \sin (3 i_0)+165 \sin (5
   i_0)+429 \sin (7 i_0)); \nonumber \\
\Delta\left.Y_2\right|_{sec} & = & - \displaystyle\frac{3}{64} A_0 \pi (3 + 5 \cos(2 i_0)) (-16 X_0 + 
   3 A_0 (3 + 5 \cos(2 i_0)) \cos(\theta_0) \nonumber \\
   & + & 14 A_0 \cos(3 \theta_0) \sin^2(i_0)); \nonumber \\
\Delta\left.i_2\right|_{sec} & = & 0; \nonumber \\
\Delta\left.\Omega_2\right|_{sec} & = & \displaystyle\frac{3}{16} A_0^2 \pi (5 \cos(3 i_0) + \cos(i_0) (7 - 60 \cos(2 \theta_0) \sin^2(i_0)))  \nonumber \\
   & + &  \frac{15}{32} \pi  A_0^2 \mathcal{J}_4 (9 \cos (i_0)+7 \cos (3 i_0))  \nonumber \\
   & - &  \frac{105}{1024} \pi  A_0^3 \mathcal{J}_6 (50 \cos (i_0)+45 \cos (3 i_0)+33 \cos (5
   i_0)).
\end{eqnarray}

\subsection{Period of the orbit}

If we apply this definition of secular variation to the time evolution of the system, we are in fact deriving the perturbed nodal period of the orbit. As such, we will represent $\Delta\left.t\right|_{sec}$ as $T$ since it is a more common denomination. That way, the zero order orbital period is:
\begin{equation}
    T_0 = 2 \pi \sqrt[4]{\frac{R^6}{A_0^3 \mu^2}}.
\end{equation}
The first order perturbation on the orbital period only depends on the effects of $J_2$, being its value:
\begin{equation}
    T_1 = - \displaystyle\frac{3}{2} A_0 \pi \sqrt[4]{\frac{R^6}{A_0^3 \mu^2}} (2 + 4 \cos(2 i_0) + 3 \cos(2 \theta_0) \sin^2(i_0)).
\end{equation}
Finally, and as previous cases, the second order effects of the zonal harmonics can be separated for the different terms of the the zonal harmonics. Particularly, the effect of $J_2$ is:
\begin{eqnarray}
    T_2|_{J_2} & = & \displaystyle\frac{3}{512} \pi \sqrt[4]{\frac{R^6}{A_0^3 \mu^2}} (-14 A_0^2 + 512 X_0^2 + 512 Y_0^2 - 824 A_0^2 \cos(2 i_0) \nonumber \\
    & + & 70 A_0^2 \cos(4 i_0) + 
    288 A_0 X_0 \cos(2 i_0 - 3 \theta_0) - 180 A_0^2 \cos(2 (i_0 - 2 \theta_0)) \nonumber \\
    & + & 
    200 A_0^2 \cos(4 i_0 - 2 \theta_0) + 32 A_0^2 \cos(2 (i_0 - \theta_0)) + 
    45 A_0^2 \cos(4 (i_0 - \theta_0)) \nonumber \\
    & + & 352 A_0 X_0 \cos(2 i_0 - \theta_0) - 192 A_0 X_0 \cos(\theta_0) - 
    464 A_0^2 \cos(2 \theta_0) \nonumber \\
    & - & 576 A_0 X_0 \cos(3 \theta_0) + 270 A_0^2 \cos(4 \theta_0) + 
    32 A_0^2 \cos(2 (i_0 + \theta_0)) \nonumber \\
    & + & 45 A_0^2 \cos(4 (i_0 + \theta_0)) + 
    352 A_0 X_0 \cos(2 i_0 + \theta_0) + 200 A_0^2 \cos(2 (2 i_0 + \theta_0)) \nonumber \\
    & - & 
    180 A_0^2 \cos(2 (i_0 + 2 \theta_0)) + 288 A_0 X_0 \cos(2 i_0 + 3 \theta_0) \nonumber \\
    & - & 
    288 A_0 Y_0 \sin(2 i_0 - 3 \theta_0) + 224 A_0 Y_0 \sin(2 i_0 - \theta_0) + 960 A_0 Y_0 \sin(\theta_0) \nonumber \\
    & - & 
    576 A_0 Y_0 \sin(3 \theta_0) - 224 A_0 Y_0 \sin(2 i_0 + \theta_0) + 288 A_0 Y_0 \sin(2 i_0 + 3 \theta_0));
\end{eqnarray}
the one of $J_3$ is:
\begin{eqnarray}
    T_2|_{J_3} & = & -\displaystyle\frac{3}{2} \pi A_0^{3/2} \mathcal{J}_3 \sqrt[4]{\frac{R^6}{A_0^3 \mu^2}}  \sin(i_0) (2 + 5 \cos(2 \theta_0) \sin^2(i_0)) \sin(\theta_0);
\end{eqnarray}
the effect of $J_4$ is:
\begin{eqnarray}
    T_2|_{J_4} & = & \displaystyle\frac{1}{256} \pi A_0^{2} \mathcal{J}_4 \sqrt[4]{\frac{R^6}{A_0^3 \mu^2}} (813 + 1540 \cos(2 i_0) + 1295 \cos(4 i_0) \nonumber \\
    & + & 240 (5 + 7 \cos(2 i_0)) \cos(2 \theta_0) \sin^2(i_0) + 840 \cos(4 \theta_0) \sin^4(i_0));
\end{eqnarray}
the perturbation from $J_5$ is:
\begin{eqnarray}
    T_2|_{J_5} & = & \displaystyle\frac{3}{128} \pi A_0^{5/2} \mathcal{J}_5 \sqrt[4]{\frac{R^6}{A_0^3 \mu^2}} \sin(i_0) \sin(
  \theta_0) (405 - 140 \cos(2 i_0) - 105 \cos(4 i_0) \nonumber \\
    & - & 2240 \sin^2(i_0) \sin^2(\theta_0) + 2016 \sin^4(i_0) \sin(\theta_0)^4);
\end{eqnarray}
the effect of $J_6$ is:
\begin{eqnarray}
    T_2|_{J_6} & = & -\displaystyle\frac{1}{16384} \pi A_0^{3} \mathcal{J}_6 \sqrt[4]{\frac{R^6}{A_0^3 \mu^2}} (56150 + 108255 \cos(2 i_0) + 95130 \cos(4 i_0) \nonumber \\
    & + & 68145 \cos(6 i_0) + 2520 (35 + 60 \cos(2 i_0) + 33 \cos(4 i_0)) \cos(2 \theta_0) \sin^2(i_0) \nonumber \\
    & + & 12096 (9 + 11 \cos(2 i_0)) \cos(4 \theta_0) \sin^4(i_0) + 44352 \cos(6 \theta_0) \sin^6(i_0));
\end{eqnarray}
and finally, the effect of $J_7$ is:
\begin{eqnarray}
    T_2|_{J_7} & = & \displaystyle\frac{1}{8192} \pi A_0^{7/2}  \mathcal{J}_7 \sqrt[4]{\frac{R^6}{A_0^3 \mu^2}} \sin(i_0) \sin(
    \theta_0) (-95270 + 23625 \cos(2 i_0) \nonumber \\
    & + & 20790 \cos(4 i_0) + 15015 \cos(6 i_0) + 967680 \sin^2(i_0) \sin^2(\theta_0) \nonumber \\
    & - & 2128896 \sin^4(i_0) \sin(\theta_0)^4 + 1317888 \sin^6(i_0) \sin(\theta_0)^6).
\end{eqnarray}

\section{Frozen condition} \label{sec:frozen}

\begin{figure}[h!]
	\centering
	{\includegraphics[width = 0.8\textwidth]{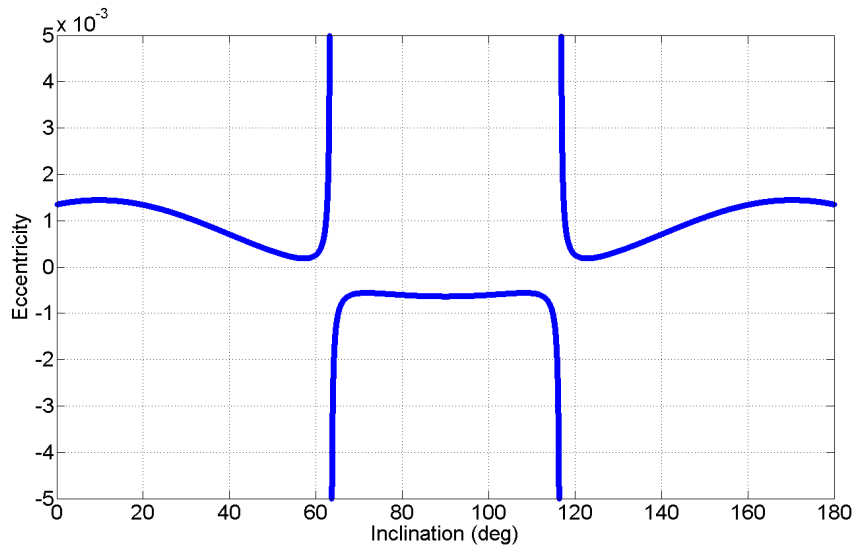}}
	\caption{Relation between initial osculating inclination and eccentricity for near circular frozen orbits.}
	\label{fig:frozen_inc_ey}
\end{figure}

The frozen condition is achieved if after one orbital revolution, the variation of $\{A,X,Y,i\}$ is equal to zero. This guarantees that the analytical solution is going to be completely periodic (but for the nodal precession of the orbit) with the argument of latitude. Therefore, we can use the results from Section~\ref{sec:secular} to impose the conditions:
\begin{eqnarray}
\Delta\left.A\right|_{sec} & = & \Delta\left.A_0\right|_{sec} + \Delta\left.A_1\right|_{sec} J_2 + \Delta\left.A_2\right|_{sec} J_2^2 = 0 \nonumber \\
\Delta\left.X\right|_{sec} & = & \Delta\left.X_1\right|_{sec} J_2 + \Delta\left.X_2\right|_{sec} J_2^2 = 0 \nonumber \\
\Delta\left.Y\right|_{sec} & = &  \Delta\left.Y_1\right|_{sec} J_2 + \Delta\left.Y_2\right|_{sec} J_2^2 = 0 \nonumber \\
\Delta\left.i\right|_{sec} & = & \Delta\left.i_0\right|_{sec} + \Delta\left.i_1\right|_{sec} J_2 + \Delta\left.i_2\right|_{sec} J_2^2 = 0.
\end{eqnarray}
This leads to the following initial condition in the components of the eccentricity vector:
\begin{eqnarray}\label{eq:frozen2}
    X_0 & = & \displaystyle\frac{1}{16} A_0 (9 \cos(\theta_0) + 15 \cos(2 i_0) \cos(\theta_0) + 14 \cos(3 \theta_0) \sin^2(i_0)); \nonumber \\
    Y_0 & = & -\frac{\sqrt{A_0}}{24576 (5 \cos (2 i_0)+3)} \big(875 A_0^2 \mathcal{J}_7 \sin (i_0)+2835 A_0^2 \mathcal{J}_7 \sin
   (3 i_0) \nonumber \\
   & + & 5775 A_0^2 \mathcal{J}_7 \sin (5 i_0)+15015 A_0^2 \mathcal{J}_7 \sin (7 i_0)-1920
   A_0\mathcal{J}_5 \sin (i_0) \nonumber \\
   & - & 6720 A_0\mathcal{J}_5 \sin (3 i_0)-20160 A_0\mathcal{J}_5 \sin
   (5 i_0)-64512 \sqrt{A_0} \sin ^2(i_0) \sin (3 \theta_0) \nonumber \\
   & - & 107520 \sqrt{A_0} \sin
   ^2(i_0) \cos (2 i_0) \sin (3 \theta_0)-119808 \sqrt{A_0} \cos (2 i_0) \sin
   (\theta_0) \nonumber \\
   & - & 80640 \sqrt{A_0} \cos (4 i_0) \sin (\theta_0)-94464 \sqrt{A_0}
   \sin (\theta_0)+36864 \mathcal{J}_3 \sin (i_0) \nonumber \\
   & + & 61440 \mathcal{J}_3 \sin (i_0) \cos
   (2 i_0)\big).
\end{eqnarray}
It is important to note that the initial value of $Y_0$ has a singularity when the initial inclination of the orbit is the critical inclination. Figure~\ref{fig:frozen_inc_ey} represents the relation introduced in Eq.~\eqref{eq:frozen2} between the initial osculating values of inclination and $y$ component of the eccentricity vector when considering an orbit with fixed $A_0 = 0.8302$ and $\theta_0 = 90$ deg. This corresponds with a set of near circular frozen orbits around the Earth with a radius of approximately at 7000 km. There are two important things to note in this figure. First, for each initial inclination there is only one possible frozen orbit with low eccentricity. Note that this is no longer true for high eccentric frozen orbits close to the critical inclination as can be seen in the following section. Second, the figure already shows the first effects on bifurcation close to the critical inclinations. This effect continues for higher eccentricities and thus, cannot be covered with the formulation used in this section. Nevertheless, the following section covers this specific case of study.

\section{Eccentric frozen orbits close to the critical inclination} \label{sec:frozen_crit}

The frozen conditions provided in Section~\ref{sec:frozen} are only applicable when the magnitude of the eccentricity of the orbit has an order of magnitude comparable with $J_2$. This situation happens in all the frozen orbits but in a small region very close to the critical inclination, where the magnitude of the eccentricity for frozen orbits increases significantly, reaching even the conditions for hyperbolic orbits. This effect was already studied by Lara, Deprit and Elipe~\cite{elipe} using a numerical continuation on families of frozen orbits. This makes the solution from Section~\ref{sec:frozen} inapplicable in this region. However, by using the series expansion from Arnas~\cite{meanj2}, it is possible not only to study this region of high eccentric orbits, but also obtain an analytical expression for the families of frozen orbits identified by Ref.~\cite{elipe}.

In case of eccentric orbits, we can no longer assume that the magnitude of the eccentricity is on the order of magnitude of the $J_2$ perturbation. Therefore, we have to work with the actual components of the eccentricity vector ($e_x, e_y$) instead of the normalized values ($X, Y$). This means that the power series expansion in $e_x$ and $e_y$ is~\cite{meanj2}:
\begin{eqnarray}
    e_x & \approx & e_{x0} + e_{x1} J_2 + e_{x2} J_2^2, \nonumber \\
    e_y & \approx & e_{y0} + e_{y1} J_2 + e_{y2} J_2^2,
\end{eqnarray}
which has to be introduced with the other variables in Eq.~\eqref{eq:difftheta} to obtain the decomposition in the power series of $J_2$. Since the effects of the term $J_2$ are covered in Arnas~\cite{meanj2,frozenj2}, in this work we just focus on the contributions of the remaining terms of the zonal harmonics. Particularly, the differential equation for the second order solution of $J_3$ is:
\begin{eqnarray}
    \displaystyle\frac{dA_2|_{J_3}}{d\theta} & = & 6 A_0^{5/2} \mathcal{J}_3 \sin (i_0) \cos (\theta)(1 + e_{x0}\cos(\theta) + e_{y0}\sin(\theta))^{2} \nonumber \\
    & \times & (5 \sin ^2(i_0) \sin
    ^2(\theta)-1); \nonumber \\
    \displaystyle\frac{de_{x2}|_{J_3}}{d\theta} & = & -A_0^{3/2} \mathcal{J}_3 (1 + e_{x0} \cos(\theta) + e_{y0} \sin(\theta))^2 (-2 \sin(i_0) \sin(
     \theta)^2 \nonumber \\
    & \times & (1 + e_{x0} \cos(\theta) + e_{y0} \sin(\theta)) (-3 + 5 \sin^2(i_0) \sin^2(\theta)) \nonumber \\
    & + & \frac{3}{2} (-1 + 5 \sin^2(i_0) \sin^2(\theta)) (e_{x0} \cos(\theta) \sin(i_0) + e_{x0} \cos^3(\theta) \sin(i_0) \nonumber \\
    & + & e_{y0} \cos(i_0) \cot(i_0) \sin(\theta) + \cos^2(\theta) \sin(i_0) (2 + e_{y0} \sin(\theta)))); \nonumber \\
    \displaystyle\frac{de_{y2}|_{J_3}}{d\theta} & = & A_0^{3/2} \mathcal{J}_3 (1 + e_{x0} \cos(\theta) + e_{y0} \sin(\theta))^2 (-2 \cos(\theta) \sin(i_0) \sin(\theta)  \nonumber \\
    & \times & (1 + e_{x0} \cos(\theta) + e_{y0} \sin(\theta)) (-3 + 5 \sin^2(i_0) \sin^2(\theta)) \nonumber \\
    & + & \frac{1}{2} (-3 + 15 \sin^2(i_0) \sin^2(\theta)) (e_{x0} \cos(i_0) \cot(i_0) \sin(\theta) \nonumber \\
    & - & \cos(\theta) \sin(i_0) (e_{y0} + (2 + e_{x0} \cos(\theta)) \sin(\theta) + e_{y0} \sin^2(\theta)))); \nonumber \\
    \displaystyle\frac{di_2|_{J_3}}{d\theta} & = & \frac{3}{2} A_0^{3/2} \mathcal{J}_3 \cos (i_0) \cos (\theta) (1 + e_{x0}\cos(\theta) + e_{y0}\sin(\theta))^{2} \nonumber \\
    & \times & (1-5 \sin ^2(i_0) \sin
   ^2(\theta)); \nonumber \\
    \displaystyle\frac{d\Omega_2|_{J_3}}{d\theta} & = & \frac{3}{2} A_0^{3/2} \mathcal{J}_3 \sin (\theta) (1 + e_{x0}\cos(\theta) + e_{y0}\sin(\theta))^{2} \nonumber \\
    & \times & (\cot (i_0)-5 \sin (i_0) \cos (i_0)
   \sin ^2(\theta));
\end{eqnarray}
for $J_4$ is:
\begin{eqnarray}
    \displaystyle\frac{dA_2|_{J_4}}{d\theta} & = & 5 A_0^3 \mathcal{J}_4 \sin ^2(i_0) \sin (2 \theta) (1 + e_{x0}\cos(\theta) + e_{y0}\sin(\theta))^{3}  \nonumber \\
    & \times & (7 \sin ^2(i_0) \sin^2(\theta)-3); \nonumber \\
    \displaystyle\frac{de_{x2}|_{J_4}}{d\theta} & = & -\frac{1}{2} A_0^2 \mathcal{J}_4 \sin(\theta) (1 + e_{x0} \cos(\theta) + e_{y0} \sin(\theta))^3 (-(1 + e_{x0} \cos(\theta)  \nonumber \\
    & + & e_{y0} \sin(\theta)) (3 - 30 \sin^2(i_0) \sin^2(\theta) + 35 \sin^4(i_0) \sin^4(\theta)) + 5 \sin(i_0)  \nonumber \\
    & \times & (-3 + 7 \sin^2(i_0) \sin^2(\theta)) (e_{x0} \cos(\theta) \sin(i_0) + e_{x0} \cos^3(\theta) \sin(i_0)  \nonumber \\
    & + & e_{y0} \cos(i_0) \cot(i_0) \sin(\theta) + \cos^2(\theta) \sin(i_0) (2 + e_{y0} \sin(\theta)))); \nonumber \\
    \displaystyle\frac{de_{y2}|_{J_4}}{d\theta} & = & \frac{1}{8} A_0^2 \mathcal{J}_4 (1 + e_{x0} \cos(\theta) + e_{y0} \sin(\theta))^3 (-5 \cos(\theta) (1 + e_{x0} \cos(\theta) \nonumber \\
    & + & e_{y0} \sin(\theta)) (3 - 30 \sin^2(i_0) \sin^2(\theta) + 35 \sin^4(i_0) \sin^4(\theta)) \nonumber \\
    & + & 20 \sin(i_0) \sin(\theta) (-3 + 7 \sin^2(i_0) \sin^2(\theta)) (e_{x0} \cos(i_0) \cot(i_0) \sin(\theta) \nonumber \\
    & - & \cos(\theta) \sin(i_0) (e_{y0} + (2 + e_{x0} \cos(\theta)) \sin(\theta) + e_{y0} \sin^2(\theta)))); \nonumber \\
    \displaystyle\frac{di_2|_{J_4}}{d\theta} & = & -\frac{5}{2} A_0^2 \mathcal{J}_4 \sin (i_0) \cos (i_0) \sin (\theta) \cos (\theta) (1 + e_{x0}\cos(\theta) + e_{y0}\sin(\theta))^{3} \nonumber \\
    & \times & (7
   \sin ^2(i_0) \sin ^2(\theta)-3); \nonumber \\
    \displaystyle\frac{d\Omega_2|_{J_3}}{d\theta} & = & -\frac{5}{2} A_0^2 \mathcal{J}_4 \cos (i_0) \sin ^2(\theta) (1 + e_{x0}\cos(\theta) + e_{y0}\sin(\theta))^{3}  \nonumber \\
    & \times & (7 \sin ^2(i_0) \sin
   ^2(\theta)-3);
\end{eqnarray}
for $J_5$ is:
\begin{eqnarray}
    \displaystyle\frac{dA_2|_{J_5}}{d\theta} & = & \frac{15}{2} A_0^{7/2} \mathcal{J}_5 \sin (i_0) \cos (\theta) (1 + e_{x0}\cos(\theta) + e_{y0}\sin(\theta))^{4} \nonumber \\
    & \times & (21 \sin ^4(i_0) \sin^4(\theta)-14 \sin ^2(i_0) \sin ^2(\theta)+1); \nonumber \\
    \displaystyle\frac{de_{x2}|_{J_5}}{d\theta} & = & -\frac{1}{8} A_0^{5/2} \mathcal{J}_5 (1 + e_{x0} \cos(\theta) + e_{y0} \sin(\theta))^4 (-4 \sin(i_0) \sin^2(\theta) (1 \nonumber \\
    & + & e_{x0} \cos(\theta) + e_{y0} \sin(\theta)) (15 - 70 \sin^2(i_0) \sin^2(\theta) + 63 \sin^4(i_0) \sin^4(\theta)) \nonumber \\
    & + & 15 (1 - 14 \sin^2(i_0) \sin^2(\theta) + 21 \sin^4(i_0) \sin^4(\theta)) (e_{x0} \cos(\theta) \sin(i_0) \nonumber \\
    & + & e_{x0} \cos^3(\theta) \sin(i_0) + e_{y0} \cos(i_0) \cot(i_0) \sin(\theta) \nonumber \\
    & + & \cos^2(\theta) \sin(i_0) (2 + e_{y0} \sin(\theta)))); \nonumber \\
    \displaystyle\frac{de_{y2}|_{J_5}}{d\theta} & = & \frac{1}{8} A_0^{5/2} \mathcal{J}_5 (1 +  e_{x0} \cos(\theta) + e_{y0} \sin(\theta))^4 (-6 \cos(\theta) \sin(i_0) \sin(\theta) (1 \nonumber \\
    & + & e_{x0} \cos(\theta) + e_{y0} \sin(\theta)) (15 - 70 \sin^2(i_0) \sin^2(\theta) + 63 \sin^4(i_0) \sin^4(\theta)) \nonumber \\
    & + & 15 (1 - 14 \sin^2(i_0) \sin^2(\theta) + 21 \sin^4(i_0) \sin^4(\theta)) (e_{x0} \cos(i_0) \cot(i_0) \sin(\theta) \nonumber \\
    & - & \cos(\theta) \sin(i_0) (e_{y0} + (2 + e_{x0} \cos(\theta)) \sin(\theta) + e_{y0} \sin^2(\theta)))); \nonumber \\
    \displaystyle\frac{di_2|_{J_5}}{d\theta} & = & -\frac{15}{8} A_0^{5/2} \mathcal{J}_5 \cos (i_0) \cos (\theta) (1 + e_{x0}\cos(\theta) + e_{y0}\sin(\theta))^{4} \nonumber \\
    & \times & (21 \sin ^4(i_0) \sin^4(\theta)-14 \sin ^2(i_0) \sin ^2(\theta)+1); \nonumber \\
    \displaystyle\frac{d\Omega_2|_{J_3}}{d\theta} & = & -\frac{15}{8} A_0^{5/2} \mathcal{J}_5 \cot (i_0) \sin (\theta) (1 + e_{x0}\cos(\theta) + e_{y0}\sin(\theta))^{4} \nonumber \\
    & \times & (21 \sin ^4(i_0) \sin
   ^4(\theta)-14 \sin ^2(i_0) \sin ^2(\theta)+1);
\end{eqnarray}
for $J_6$ is:
\begin{eqnarray}
    \displaystyle\frac{dA_2|_{J_6}}{d\theta} & = & \frac{21}{2} A_0^4 \mathcal{J}_6 \sin ^2(i_0) \sin (\theta) \cos (\theta) (1 + e_{x0}\cos(\theta) + e_{y0}\sin(\theta))^{5} \nonumber \\
    & \times & (33 \sin^4(i_0) \sin ^4(\theta)-30 \sin ^2(i_0) \sin ^2(\theta)+5); \nonumber \\
    \displaystyle\frac{de_{x2}|_{J_6}}{d\theta} & = & -\frac{1}{16} A_0^3 \mathcal{J}_6 (1 + e_{x0} \cos(\theta) + e_{y0} \sin(\theta))^5 (-4 \sin(\theta) (1 + e_{x0} \cos(\theta) \nonumber \\
    & + & e_{y0} \sin(\theta)) (-5 + 105 \sin^2(i_0) \sin^2(\theta) - 315 \sin^4(i_0) \sin^4(\theta) \nonumber \\
    & + & 231 \sin^6(i_0) \sin^6(\theta)) + 42 \sin(i_0) \sin(\theta) (5 - 30 \sin^2(i_0) \sin^2(\theta) \nonumber \\
    & + & 33 \sin^4(i_0) \sin^4(\theta)) (e_{x0} \cos(\theta) \sin(i_0) + e_{x0} \cos^3(\theta) \sin(i_0) \nonumber \\
    & + & e_{y0} \cos(i_0) \cot(i_0) \sin(\theta) + \cos^2(\theta) \sin(i_0) (2 + e_{y0} \sin(\theta)))); \nonumber \\
    \displaystyle\frac{de_{y2}|_{J_6}}{d\theta} & = & \frac{1}{16} A_0^3 \mathcal{J}_6 (1 + e_{x0} \cos(\theta) + e_{y0} \sin(\theta))^5 (-7 \cos(\theta) (1 + e_{x0} \cos(\theta) \nonumber \\
    & + & e_{y0} \sin(\theta)) (-5 + 105 \sin^2(i_0) \sin^2(\theta) - 315 \sin^4(i_0) \sin^4(\theta) \nonumber \\
    & + & 231 \sin^6(i_0) \sin^6(\theta)) + 42 \sin(i_0) \sin(\theta) (5 - 30 \sin^2(i_0) \sin^2(\theta) \nonumber \\
    & + & 33 \sin^4(i_0) \sin^4(\theta)) (e_{x0} \cos(i_0) \cot(i_0) \sin(\theta) - \cos(\theta) \sin(i_0) (e_{y0} \nonumber \\
    & + & (2 + e_{x0} \cos(\theta)) \sin(\theta) + e_{y0} \sin^2(\theta)))); \nonumber \\
    \displaystyle\frac{di_2|_{J_6}}{d\theta} & = & -\frac{21}{8} A_0^3 \mathcal{J}_6 \sin (i_0) \cos (i_0) \sin (\theta) \cos (\theta) (1 + e_{x0}\cos(\theta) + e_{y0}\sin(\theta))^{5} \nonumber \\
    & \times & (33
   \sin ^4(i_0) \sin ^4(\theta) - 30 \sin ^2(i_0) \sin ^2(\theta)+5); \nonumber \\
    \displaystyle\frac{d\Omega_2|_{J_3}}{d\theta} & = & -\frac{21}{8} A_0^3 \mathcal{J}_6 \cos (i_0) \sin ^2(\theta) (1 + e_{x0}\cos(\theta) + e_{y0}\sin(\theta))^{5} \nonumber \\
    & \times & (33 \sin ^4(i_0) \sin
   ^4(\theta)-30 \sin ^2(i_0) \sin ^2(\theta)+5);
\end{eqnarray}
and for $J_7$ is:
\begin{eqnarray}
    \displaystyle\frac{dA_2|_{J_7}}{d\theta} & = & \frac{7}{4} A_0^{9/2} \mathcal{J}_7 \sin (i_0) \cos (\theta) (1 + e_{x0}\cos(\theta) + e_{y0}\sin(\theta))^{6} \nonumber \\
    & \times & (429 \sin ^6(i_0) \sin^6(\theta)-495 \sin ^4(i_0) \sin ^4(\theta) + 135 \sin ^2(i_0) \sin ^2(\theta)-5); \nonumber \\
    \displaystyle\frac{de_{x2}|_{J_7}}{d\theta} & = & -\frac{1}{16} A_0^{7/2} \mathcal{J}_7 (1 + e_{x0} \cos(\theta) + e_{y0} \sin(\theta))^6 (-4 \sin(i_0) \sin^2(\theta) \nonumber \\
    & \times & (1 + e_{x0} \cos(\theta) + e_{y0} \sin(\theta)) (-35 + 315 \sin^2(i_0) \sin^2(\theta) \nonumber \\
    & - & 693 \sin^4(i_0) \sin^4(\theta) + 429 \sin^6(i_0) \sin^6(\theta)) + 
   7 (-5 + 135 \sin^2(i_0) \sin^2(\theta) \nonumber \\
    & - & 495 \sin^4(i_0) \sin^4(\theta) + 429 \sin^6(i_0) \sin^6(\theta)) (e_{x0} \cos(\theta) \sin(i_0) \nonumber \\
    & + & e_{x0} \cos^3(\theta) \sin(i_0) + e_{y0} \cos(i_0) \cot(i_0) \sin(\theta) \nonumber \\
    & + & \cos^2(\theta) \sin(i_0) (2 + e_{y0} \sin(\theta)))); \nonumber \\
    \displaystyle\frac{de_{y2}|_{J_7}}{d\theta} & = & \frac{1}{16} A_0^{7/2} \mathcal{J}_7 (1 + e_{x0} \cos(\theta) + e_{y0} \sin(\theta))^6 (-8 \cos(\theta) \sin(i_0) \sin(\theta) \nonumber \\
    & \times & (1 + e_{x0} \cos(\theta) + e_{y0} \sin(\theta)) (-35 + 315 \sin^2(i_0) \sin^2(\theta) \nonumber \\
    & - & 693 \sin^4(i_0) \sin^4(\theta) + 429 \sin^6(i_0) \sin^6(\theta)) + 7 (-5 + 135 \sin^2(i_0) \sin^2(\theta) \nonumber \\
    & - & 495 \sin^4(i_0) \sin^4(\theta) + 429 \sin^6(i_0) \sin^6(\theta)) (e_{x0} \cos(i_0) \cot(i_0) \sin(\theta) \nonumber \\
    & - & \cos(\theta) \sin(i_0) (e_{y0} + (2 + e_{x0} \cos(\theta)) \sin(\theta) + e_{y0} \sin^2(\theta)))); \nonumber \\
    \displaystyle\frac{di_2|_{J_7}}{d\theta} & = & -\frac{7}{16} A_0^{7/2} \mathcal{J}_7 \cos (i_0) \cos (\theta) (1 + e_{x0}\cos(\theta) + e_{y0}\sin(\theta))^{6} \nonumber \\
    & \times &  (429 \sin ^6(i_0) \sin ^6(\theta) - 495 \sin ^4(i_0) \sin ^4(\theta)+135 \sin ^2(i_0) \sin ^2(\theta)-5); \nonumber \\
    \displaystyle\frac{d\Omega_2|_{J_7}}{d\theta} & = & -\frac{7}{16} A_0^{7/2} \mathcal{J}_7 \cot (i_0) \sin (\theta) (1 + e_{x0}\cos(\theta) + e_{y0}\sin(\theta))^{6} \nonumber \\
    & \times &  (429 \sin ^6(i_0)
   \sin ^6(\theta) - 495 \sin ^4(i_0) \sin ^4(\theta)+135 \sin ^2(i_0) \sin ^2(\theta)-5).
\end{eqnarray}

The solution to these systems of equations becomes quite large, more so if we want to include the effects of all five terms of the zonal harmonics. Therefore, the complete analytical expressions are not included in this manuscript, but they can be accessed as Matlab scripts in the following web page: \\
\href{https://engineering.purdue.edu/ART/research/research-code}{https://engineering.purdue.edu/ART/research/research-code}. \\ In here, instead, we focus on the secular variation of the orbital elements, as they are the ones defining the conditions for frozen orbits. Particularly, the second order secular variation on variable $A$ after one orbital revolution is:
\begin{eqnarray}  
    \Delta\left.A_2\right|_{J_2sec} & = & -3 A_0^3 \pi e_{x0} e_{y0} \sin^2(i_0); \nonumber \\
    \Delta\left.A_2\right|_{J_3sec} & = & -\frac{3}{2} \pi  A_0^{5/2} \mathcal{J}_3 e_{x0} \sin (i_0) (5 \cos (2 i_0)+3); \nonumber \\
    \Delta\left.A_2\right|_{J_4sec} & = & -\frac{15}{4} \pi  A_0^3 \mathcal{J}_4 e_{x0} e_{y0} \sin ^2(i_0) (7 \cos (2 i_0)+5); \nonumber \\
    \Delta\left.A_2\right|_{J_5sec} & = & \frac{15}{256} \pi  A_0^{7/2} \mathcal{J}_5 e_{x0} \sin (i_0) \big(28 \cos (2 i_0)
   \big(7 e_{x0}^2+3 e_{y0}^2+8\big) \nonumber \\
    & + & 21 \cos (4 i_0) \left(3 e_{x0}^2+15 e_{y0}^2+8\right)+5
   \left(25 e_{x0}^2-3 e_{y0}^2+24\right)\big); \nonumber \\
    \Delta\left.A_2\right|_{J_6sec} & = & \frac{105}{512} \pi  A_0^4 \mathcal{J}_6 e_{x0} e_{y0} \sin ^2(i_0) \big(12 \cos (2 i_0)
   \left(27 e_{x0}^2+23 e_{y0}^2+50\right) \nonumber \\
    & + & 33 \cos (4 i_0) \left(3 e_{x0}^2+7
   e_{y0}^2+10\right)+7 \left(31 e_{x0}^2+19 e_{y0}^2+50\right)\big); \nonumber \\
    \Delta\left.A_2\right|_{J_7sec} & = & -\frac{105}{32768} \pi  A_0^{9/2} \mathcal{J}_7 e_{x0} \sin (i_0) \big(429 e_{x0}^4 \cos (6 i_0)+6006
   e_{x0}^2 e_{y0}^2 \cos (6 i_0) \nonumber \\
    & + & 3432 e_{x0}^2 \cos (6 i_0)+9 \cos (2 i_0) \big(523 e_{x0}^4+10
   e_{x0}^2 \left(41 e_{y0}^2+188\right)-25 e_{y0}^4 \nonumber \\
    & + & 360 e_{y0}^2+600\big)+198 \cos (4 i_0)
   \big(11 e_{x0}^4+e_{x0}^2 \left(58 e_{y0}^2+56\right)+7 e_{y0}^4 \nonumber \\
    & + & 72 e_{y0}^2+24\big)+9009 e_{y0}^4
   \cos (6 i_0)+24024 e_{y0}^2 \cos (6 i_0) + 3432 \cos (6 i_0) \nonumber \\
    & + & +2926 e_{x0}^4-700 e_{x0}^2
   e_{y0}^2+9520 e_{x0}^2+70 e_{y0}^4-560 e_{y0}^2+2800\big);
\end{eqnarray}
where it is important to note that all the terms of the gravitational potential have as a common divisor $e_{x0}\sin(i_0)$. The same effect happens with the secular variation in the inclination of the orbit. This means that in order to make these secular variations zero for the purposes of a second order solution, the only possibilities available are either $\mathcal{O}(e_{x0}) = \mathcal{O}(J_2)$ or $\mathcal{O}(i_{0}) = \mathcal{O}(J_2)$. However, close to equatorial orbits cannot banish the secular variation of the components of the eccentricity vector (see for instance Eq.~\eqref{eq:jnexsec}). Therefore, the only real possibility to cancel the secular effects in the orbital elements $A$, $i$, $e_x$, and $e_y$ to generate a frozen orbit is $\mathcal{O}(e_{x0}) = \mathcal{O}(J_2)$. In fact, the condition $\mathcal{O}(e_{x0}) = \mathcal{O}(J_2)$ already makes the orbital elements $A$ and $i$ periodic under a second order solution.

Therefore, and under the condition $\mathcal{O}(e_{x0}) = \mathcal{O}(J_2)$, the first order secular effects on the components of the eccentricity vector $e_x$ and $e_y$ become:
\begin{eqnarray}  
    \Delta\left.e_{x1}\right|_{sec} & = &  -\frac{3}{4} A_0 \pi e_{y0} (3 + 5 \cos(2 i_0)); \nonumber \\
    \Delta\left.e_{y1}\right|_{sec} & = & 0.
\end{eqnarray}
On the other hand, the second order secular variation in $e_x$ can be separated in the different terms of the zonal harmonics.
\begin{eqnarray}  \label{eq:jnexsec}
    \Delta\left.e_{x2}\right|_{J_2sec} & = & -\frac{3}{1024} \pi  A_0^2 \big(-36 e_{y0}^2 \sin (2 i_0-5 \theta_0)+45 e_{y0}^2 \sin (4 i_0-5
   \theta_0) \nonumber \\
    & - & 380 e_{y0}^2 \sin (2 i_0-3 \theta_0)+55 e_{y0}^2 \sin (4 i_0-3
   \theta_0)+1320 e_{y0}^2 \sin (2 i_0-\theta_0) \nonumber \\
    & - & 390 e_{y0}^2 \sin (4
   i_0-\theta_0)-1320 e_{y0}^2 \sin (2 i_0+\theta_0)+390 e_{y0}^2 \sin (4
   i_0+\theta_0) \nonumber \\
    & + & 380 e_{y0}^2 \sin (2 i_0+3 \theta_0)-55 e_{y0}^2 \sin (4 i_0+3
   \theta_0)+36 e_{y0}^2 \sin (2 i_0+5 \theta_0) \nonumber \\
    & - & 45 e_{y0}^2 \sin (4 i_0+5
   \theta_0)+144 e_{y0} \cos (2 (i_0-2 \theta_0))-360 e_{y0} \cos (4 i_0-2
   \theta_0) \nonumber \\
    & + & 1056 e_{y0} \cos (2 (i_0-\theta_0))-180 e_{y0} \cos (4
   (i_0-\theta_0)) \nonumber \\
    & + & 1056 e_{y0} \cos (2 (i_0+\theta_0))-180 e_{y0} \cos (4
   (i_0+\theta_0)) \nonumber \\
    & - & 360 e_{y0} \cos (2 (2 i_0+\theta_0))+144 e_{y0} \cos (2 (i_0+2
   \theta_0))+112 \sin (2 i_0-3 \theta_0) \nonumber \\
    & - & 140 \sin (4 i_0-3 \theta_0)+624
   \sin (2 i_0-\theta_0)+420 \sin (4 i_0-\theta_0) \nonumber \\
    & - & 624 \sin (2
   i_0+\theta_0)-420 \sin (4 i_0+\theta_0)-112 \sin (2 i_0+3 \theta_0) \nonumber \\
    & + & 140
   \sin (4 i_0+3 \theta_0)+280 e_{y0}^3 \cos (2 i_0)-630 e_{y0}^3 \cos (4 i_0) \nonumber \\
    & + & 592 e_{y0}
   \cos (2 i_0)-820 e_{y0} \cos (4 i_0)-1980 e_{y0}^2 \sin (\theta_0)+630 e_{y0}^2 \sin (3
   \theta_0) \nonumber \\
    & + & 18 e_{y0}^2 \sin (5 \theta_0)+1680 e_{y0} \cos (2 \theta_0)+72 e_{y0}
   \cos (4 \theta_0)-984 \sin (\theta_0) \nonumber \\
    & - & 56 \sin (3 \theta_0)+350
   e_{y0}^3+228 e_{y0}\big); \nonumber \\
    \Delta\left.e_{x2}\right|_{J_3sec} & = & \frac{3}{32} A_0^{3/2} \mathcal{J}_3 \pi (-1 - 3 e_{y0}^2 - 4 \cos(2 i_0) + 5 (1 + 7 e_{y0}^2) \cos(4 i_0)) \csc(i_0); \nonumber \\
    \Delta\left.e_{x2}\right|_{J_4sec} & = & \frac{1}{1024} A_0^2 \mathcal{J}_4 \pi e_{y0} (33 (88 + 131 e_{y0}^2) + 20 (280 + 467 e_{y0}^2) \cos(2 i_0) \nonumber \\
    & + & 35 (152 + 235 e_{y0}^2) \cos(4 i_0))
; \nonumber \\
    \Delta\left.e_{x2}\right|_{J_5sec} & = & \frac{5}{8192} A_0^{5/2} \mathcal{J}_5 \pi (-3072 e_{y0}^2 (4 + 3 e_{y0}^2) \csc(i_0) + 6 (16 + 4328 e_{y0}^2 \nonumber \\
    & + & 3037 e_{y0}^4) \sin(i_0) + 7 (48 + 3896 e_{y0}^2 + 2735 e_{y0}^4) \sin(3 i_0) + 63 (16 \nonumber \\
    & + & 360 e_{y0}^2 + 245 e_{y0}^4) \sin(5 i_0))
; \nonumber \\
    \Delta\left.e_{x2}\right|_{J_6sec} & = & -\frac{3}{131072} A_0^3 \mathcal{J}_6 \pi e_{y0} (141200 + 749200 e_{y0}^2 + 303570 e_{y0}^4 + 35 (7864 \nonumber \\
    & + & 44120 e_{y0}^2 + 17151 e_{y0}^4) \cos(2 i_0) + 42 (6040 + 38040 e_{y0}^2 + 14819 e_{y0}^4) \nonumber \\
    & \times & \cos(4 i_0) + 231000 \cos(6 i_0) + 1228920 e_{y0}^2 \cos(6 i_0) + 470547 e_{y0}^4 \cos(6 i_0))
; \nonumber \\
    \Delta\left.e_{x2}\right|_{J_7sec} & = & \frac{7}{524288} A_0^{7/2} \mathcal{J}_7 \pi (-1000 - 33720 e_{y0}^2 + 6775 e_{y0}^4 - 1400 e_{y0}^6 \nonumber \\
    & + & 4 (-560 - 10968 e_{y0}^2 - 8530 e_{y0}^4 + 1049 e_{y0}^6) \cos(2 i_0) - 12 (280 \nonumber \\
    & - & 3048 e_{y0}^2 + 4575 e_{y0}^4 + 1470 e_{y0}^6) \cos(4 i_0) - 10560 \cos(6 i_0) \nonumber \\
    & + & 231264 e_{y0}^2 \cos(6 i_0) + 693000 e_{y0}^4 \cos(6 i_0) \nonumber \\
    & + & 176220 e_{y0}^6 \cos(6 i_0) + 17160 \cos(8 i_0) + 792792 e_{y0}^2 \cos(8 i_0) \nonumber \\
    & + & 1846845 e_{y0}^4 \cos(8 i_0) + 453024 e_{y0}^6 \cos(8 i_0)) \csc(i_0).
\end{eqnarray}
Finally, the second order secular variation of $e_y$ only depends on the effects of $J_2$ under these conditions:
\begin{eqnarray}
    \Delta\left.e_{y2}\right|_{J_2sec} & = & -\frac{9}{32} A_0^2 \pi^2 e_{y0} (3 + 5 \cos(2 i_0))^2.
\end{eqnarray}
This secular variation in $e_y$ can only banish in a second order solution if either $\mathcal{O}(e_{y0}) = \mathcal{O}(J_2)$ or if $\mathcal{O}(3 + 5 \cos(2 i_0)) = \mathcal{O}(J_2)$. The first possibility corresponds to the low eccentric orbits already studied in Section~\ref{sec:frozen} so we do not repeat their analysis in here. Conversely, the condition $\mathcal{O}(3 + 5 \cos(2 i_0)) = \mathcal{O}(J_2)$ corresponds to frozen orbits close to the critical inclination that can have arbitrary large values of the component $e_y$ of the eccentricity vector.

Therefore, it is possible to obtain the frozen condition in this region close to the critical inclination just by analyzing the secular variation of $e_x$ up to second order. In particular, we can impose that:
\begin{eqnarray}
    \Delta\left.e_{x}\right|_{sec} & = & \Delta\left.e_{x1}\right|_{sec} + J_2\Big(\Delta\left.e_{x2}\right|_{J_2sec} + \Delta\left.e_{x2}\right|_{J_3sec} + \Delta\left.e_{x2}\right|_{J_4sec} \nonumber \\
    & + & \Delta\left.e_{x2}\right|_{J_5sec} + \Delta\left.e_{x2}\right|_{J_6sec} + \Delta\left.e_{x2}\right|_{J_7sec}\Big) = 0.
\end{eqnarray}
However, since the frozen orbit happens in the condition $\mathcal{O}(3 + 5 \cos(2 i_0)) = \mathcal{O}(J_2)$, we can perform the following change of variable for the inclination:
\begin{equation} \label{eq:critexp}
    K J_2 = 3 + 5 \cos(2 i_0),
\end{equation}
and keep all the terms of order $J_2^2$ or larger. This leads to the following condition relating $K$, $A_0$, $\theta_0$, and $e_{y0}$:
\begin{eqnarray} \label{eq:frozeney}
    \Delta\left.e_{x}\right|_{sec} & = & -\frac{3}{4} \pi J_2^2 A_0
   K e_{y0}+ \frac{1}{32000} \pi J_2^2 A_0^{3/2} \Bigg(-4800 \sqrt{A_0} e_{y0} \big(-12 e_{y0} \sin
   (\theta_0) \nonumber \\
    & + & 4 e_{y0} \sin (3 \theta_0)+12 \cos (2 \theta_0)+7
   e_{y0}^2+2\big)  \nonumber \\
    & - & 38400 \mathcal{J}_3 e_{y0}^2 \sec \left(\frac{1}{2} \arccos \left(\frac{3}{5}\right)\right) - \sqrt{A_0} \mathcal{J}_4 (112000 
   e_{y0}^3-60800 e_{y0}) \nonumber \\
    & - & 160 A_0 \mathcal{J}_5 \left(865 e_{y0}^4+1180
   e_{y0}^2+72\right) \sec \left(\frac{1}{2} \arccos \left(\frac{3}{5}\right)\right)  \nonumber \\
    & - & 6 A_0^{3/2} \mathcal{J}_6 \left(25581 e_{y0}^4+64160
   e_{y0}^2+14800\right) e_{y0} \nonumber \\
    & - &  7 A_0^2 \mathcal{J}_7 \left(32963 e_{y0}^6+132140 e_{y0}^4+71664
   e_{y0}^2+3520\right)  \nonumber \\
    & \times & \sin \left(\frac{1}{2} \arccos \left(\frac{3}{5}\right)\right)\Bigg) + \mathcal{O}(J_2^3) .
\end{eqnarray}
This expression is a sixth order equation in $e_{y0}$ and does not have, in general, analytical solution. Nevertheless, since the expression is analytic, it is possible to obtain its roots using a numerical approach such as a Housholder's method or a bracketing method. However, there is a more simple approach to find the frozen condition assuming that the initial value $e_{y0}$ is known. From the previous expression, it is possible to obtain a closed form solution for $K$ that is unique for any combination of $A_0$, $\theta_0$, and $e_{y0}$. Particularly:
\begin{eqnarray} \label{eq:kvalue}
    K & = & \displaystyle\frac{12}{5} A_0 e_{y0} \sin
   (\theta_0)-\frac{4}{5} A_0 e_{y0} \sin (3 \theta_0)-\frac{12}{5} A_0 \cos (2
   \theta_0)-\frac{7 A_0 e_{y0}^2}{5}-\frac{2 A_0}{5} \nonumber \\
    & - & 4 \sqrt{A_0} \mathcal{J}_3 e_{y0} \sin
   \left(\frac{1}{2} \arccos \left(\frac{3}{5}\right)\right)-\frac{14}{3} A_0 \mathcal{J}_4
   e_{y0}^2-\frac{38 A_0 \mathcal{J}_4}{15} \nonumber \\
    & - & \frac{173}{30} A_0^{3/2} \mathcal{J}_5 e_{y0}^3 \sec \left(\frac{1}{2} \arccos \left(\frac{3}{5}\right)\right)-\frac{59}{3} A_0^{3/2} \mathcal{J}_5 e_{y0}
   \sin \left(\frac{1}{2} \arccos \left(\frac{3}{5}\right)\right) \nonumber \\
    & - & \frac{6 A_0^{3/2}}{5
   e_{y0}} \mathcal{J}_5 \sin
   \left(\frac{1}{2} \arccos \left(\frac{3}{5}\right)\right)-\frac{25581}{4000} A_0^2 \mathcal{J}_6 e_{y0}^4-\frac{401}{25} A_0^2 \mathcal{J}_6
   e_{y0}^2-\frac{37}{10} A_0^2 \mathcal{J}_6 \nonumber \\
    & - & \frac{230741}{24000} A_0^{5/2} \mathcal{J}_7 e_{y0}^5 \sin \left(\frac{1}{2} \arccos \left(\frac{3}{5}\right)\right)-\frac{46249}{1200} A_0^{5/2}
   \mathcal{J}_7 e_{y0}^3 \sin \left(\frac{1}{2} \arccos \left(\frac{3}{5}\right)\right) \nonumber \\
    & - & \frac{10451}{500} A_0^{5/2}
   \mathcal{J}_7 e_{y0} \sin \left(\frac{1}{2} \arccos \left(\frac{3}{5}\right)\right)-\frac{77 A_0^{5/2}}{75
   e_{y0}} \mathcal{J}_7 \sin
   \left(\frac{1}{2} \arccos \left(\frac{3}{5}\right)\right).
\end{eqnarray}
This solution can be then introduced in Eq.~\eqref{eq:critexp} to obtain the initial value of the inclination. In here, it is important to note two things. First, the value of $K$ becomes singular if $e_{y0}$ tends to zero. This case corresponds with near circular frozen orbits and as such, can be evaluated using the approach presented in Section~\ref{sec:frozen}. Second, there is no specific value given to $e_{x0}$, just the constraint that $\mathcal{O}(e_{x0}) = \mathcal{O}(J_2)$. This means that frozen orbits close to the critical inclination are frozen under a second order approximation as long as the $x$ component of the eccentricity vector is in the order of magnitude of $J_2$ and the condition in $e_{y0}$ (Eq.~\eqref{eq:frozeney}) is satisfied.

\begin{figure}[h!]
	\centering
	{\includegraphics[width = 0.8\textwidth]{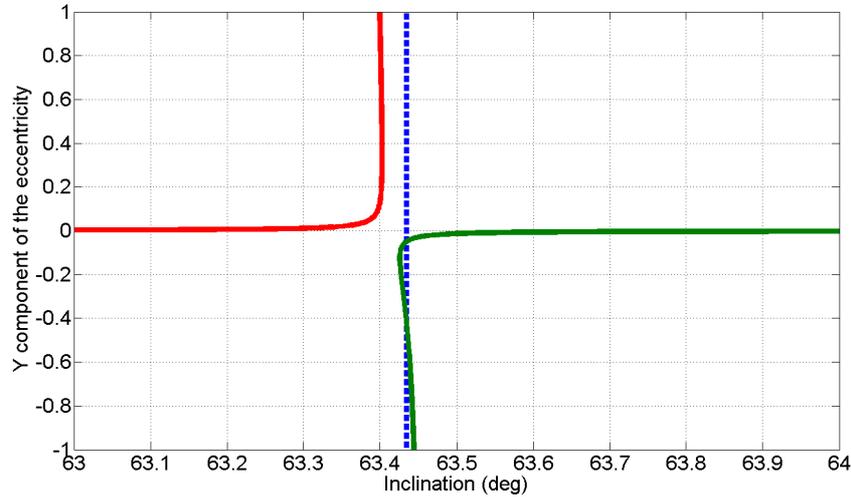}}
	\caption{Relation between initial osculating inclination and eccentricity for frozen orbits close to the critical inclination.}
	\label{fig:bifurcation_osc}
\end{figure}

Figure~\ref{fig:bifurcation_osc} shows the relation between the osculating inclination and $e_{y0}$ as an example of application of these expressions. In this figure, we assumed an initial position at $\theta_0 = 90$ deg and orbits whose periapsis about the Earth has a value of 650 km of altitude. The left curve corresponds to orbits with a value of $e_{y0}$ that is positive, that is, the periapsis is over the Norther hemisphere. Conversely, the curve on the right corresponds to orbits with $e_{y0 < 0}$, while the dashed line represents the position of the critical inclination. As it happened with the near circular frozen orbits, bifurcation can be seen close to the critical inclination.

\begin{figure}[h!]
	\centering
	{\includegraphics[width = 0.8\textwidth]{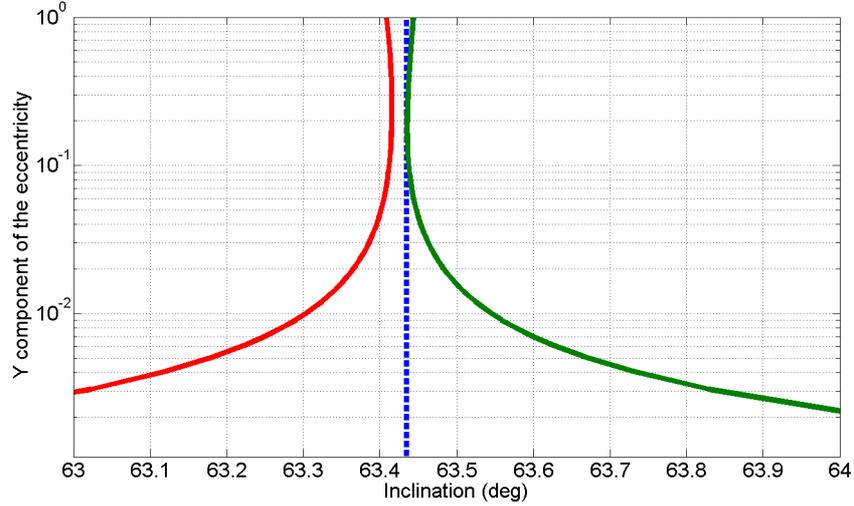}}
	\caption{Relation between initial mean inclination and eccentricity for frozen orbits close to the critical inclination.}
	\label{fig:bifurcation_mean}
\end{figure}

This effect can also be observed if we transform these osculating values to their mean values using the approach from Section~\ref{sec:mean}. The result of this transformation is included in Fig.~\ref{fig:bifurcation_mean}, where the bifurcation in the vicinity of the critical inclination can be observed. Note that the critical inclination separates both families of frozen orbits, but not symmetrically. In fact, there are several ways in which this morphology can be changed. 

\begin{figure}[h!]
	\centering
	{\includegraphics[width = 0.48\textwidth]{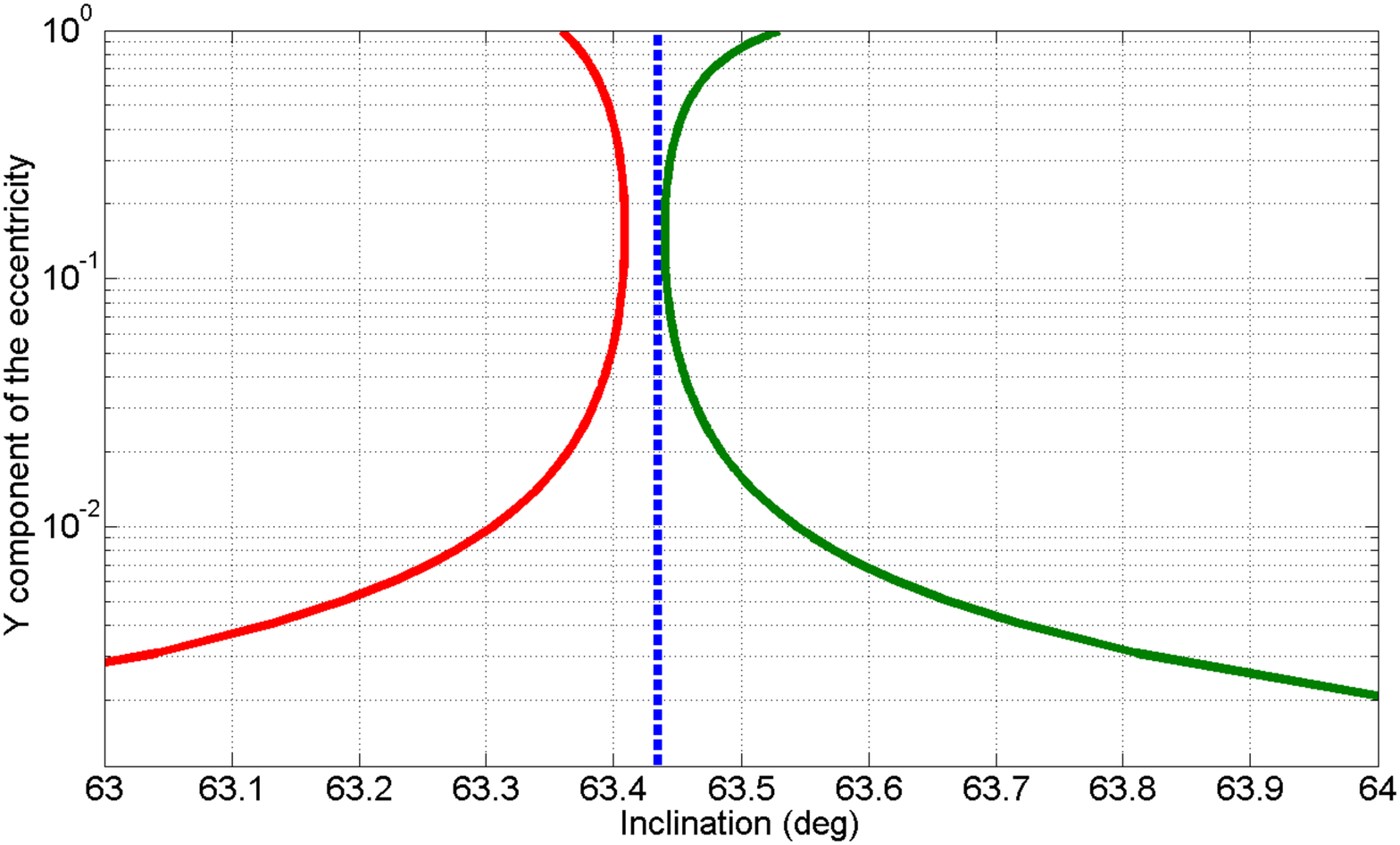}} \hfill {\includegraphics[width = 0.48\textwidth]{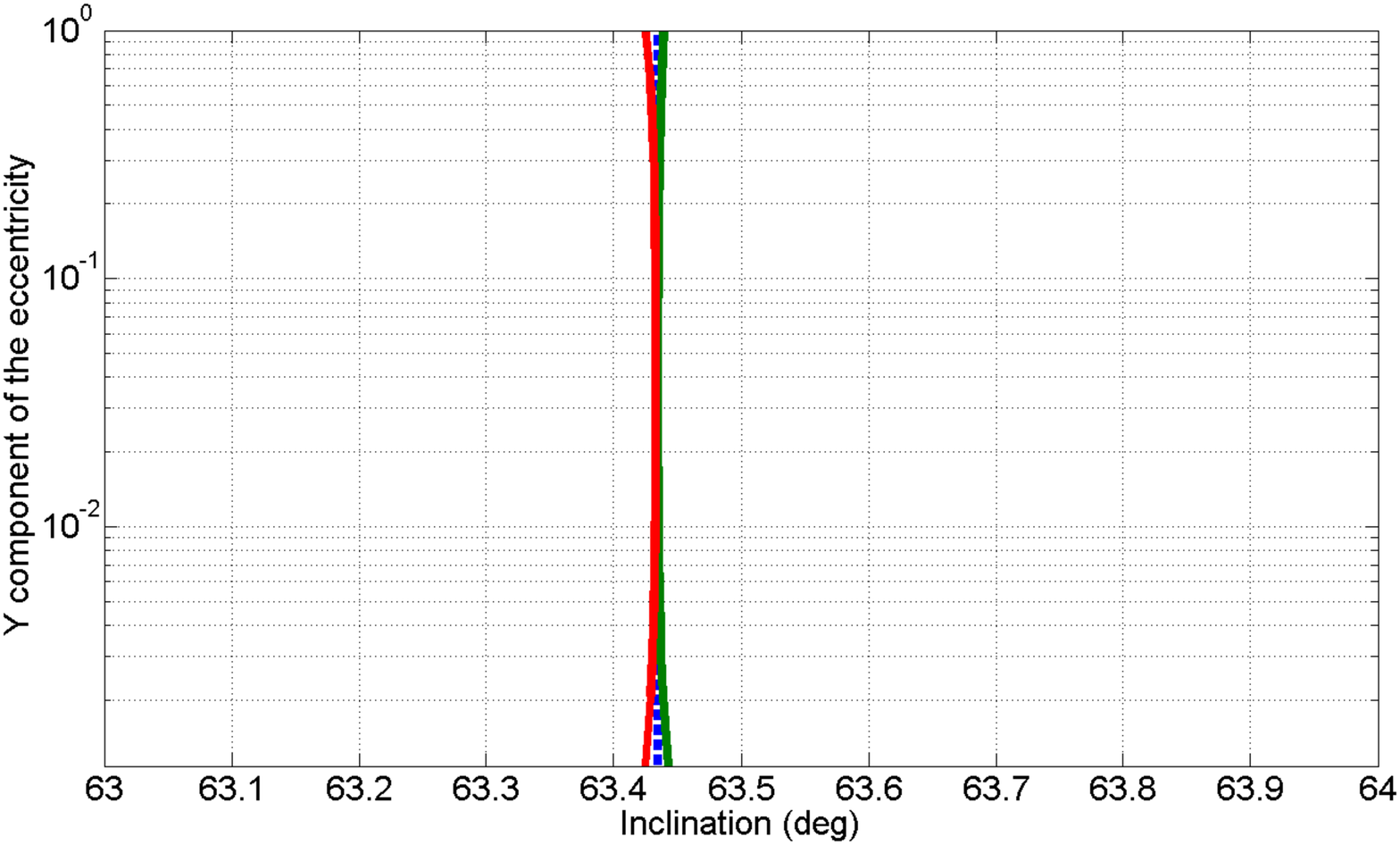}}
	\caption{Effect of changing the value of $A_0$ in the mean distribution of eccentricity and inclination in frozen orbits close to the critical inclination.}
	\label{fig:bifurcation2}
\end{figure}

For instance, if instead of fixing the altitude of the periapsis, the value of $A_0$ is fixed, we can obtain the frozen families seen in Fig.~\ref{fig:bifurcation2}. The left plot corresponds with $A_0 = 0.8$ while the right one with $A_0 = 0.05$. These two curves tend to the critical inclination line as the value of $A_0$ decreases as can be already analyzed in Eq.~\eqref{eq:kvalue}. Conversely, as the value of $A_0$ increases, the curves get further away from the critical inclination line, and additionally, their upper extremes (the ones related with higher values of the eccentricity) curve more towards the exterior of the critical inclination. Note also that some of these orbits are unrealistic from a practical point of view since they represent orbits in collision with the Earth's surface.


\section{Examples of application} \label{sec:examples}

In this section, several examples of application of the proposed methodologies are presented in order to show the their performance in defining and studying frozen orbits at different initial conditions. Particularly, we present an example of near circular frozen orbit and a second one dealing with an eccentric frozen orbit close to the critical inclination. All the examples consider the Earth as the main celestial body. 

\subsection{Near circular frozen orbit}

\begin{figure}[h!]
	\centering
	{\includegraphics[width = 0.95\textwidth]{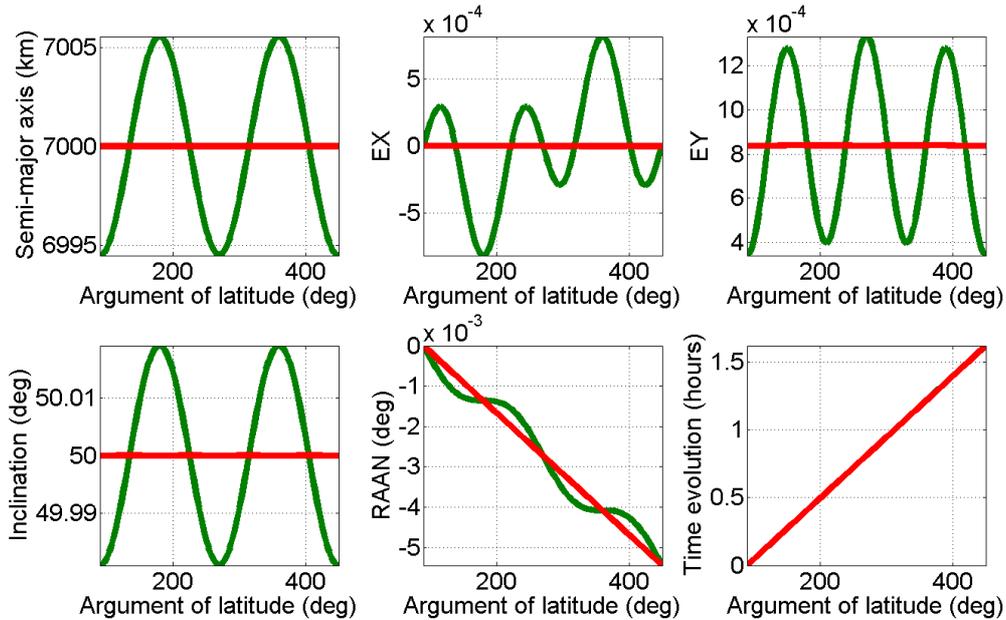}}
	\caption{Osculating and mean evolution of a near circular frozen orbit.}
	\label{fig:mean_50deg}
\end{figure}

In this example of application, a near circular frozen orbit at a mean semi-major axis of 7000 km with a mean inclination of 50 deg is selected. Particularly, from the initial conditions: $A_0 = 0.8315$, $i_0 = 49.981$ deg, $\theta_0 = 90$ deg, the initial value of the two components of the eccentricity vector can be derived using Eq.~\eqref{eq:frozen2}. This leads to $e_{x0} = 0$, and $e_{y0} = 3.3882\cdot 10^{-4}$. Additionally, $\Omega_0$ as well as the initial time has been selected to start at zero. Figure~\ref{fig:mean_50deg} shows the evolution of these orbital elements as a function of the argument of latitude, as well as their respective mean value. In this regard, the mean value has been obtained combining the transformations provided in Section~\ref{sec:mean} and Ref.~\cite{frozenj2} applied to the numerical solution in each time step. As can be seen, the orbital elements semi-major axis, inclination and the two components of the eccentricity vector are completely periodic. This effect can also be seen in the fact the the mean value of these orbital elements is also maintained over time.

\begin{figure}[h!]
	\centering
	{\includegraphics[width = 0.95\textwidth]{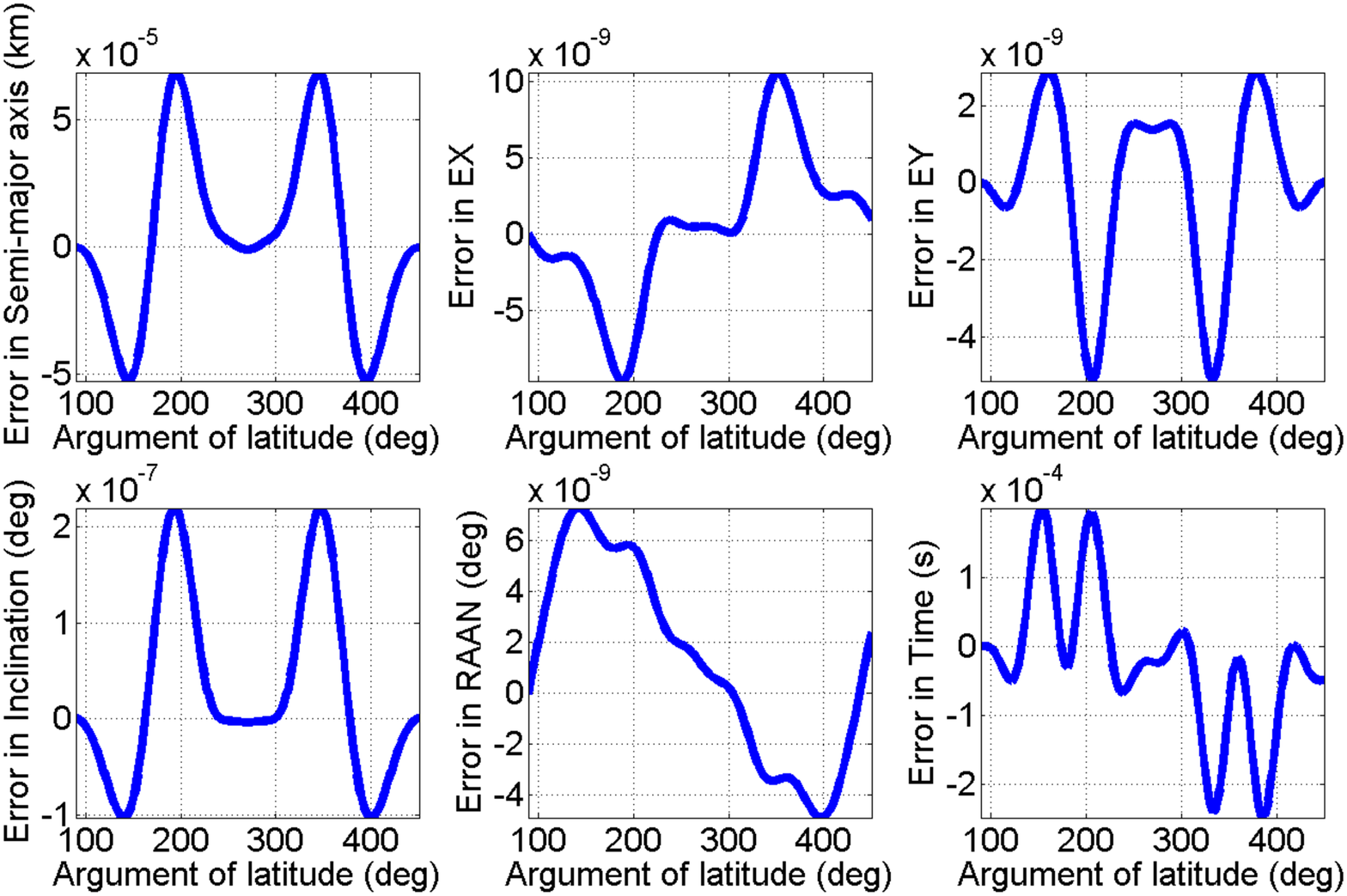}}
	\caption{Second order solution error of a near circular frozen orbit.}
	\label{fig:error_50deg}
\end{figure}

\begin{figure}[h!]
	\centering
	{\includegraphics[width = 0.95\textwidth]{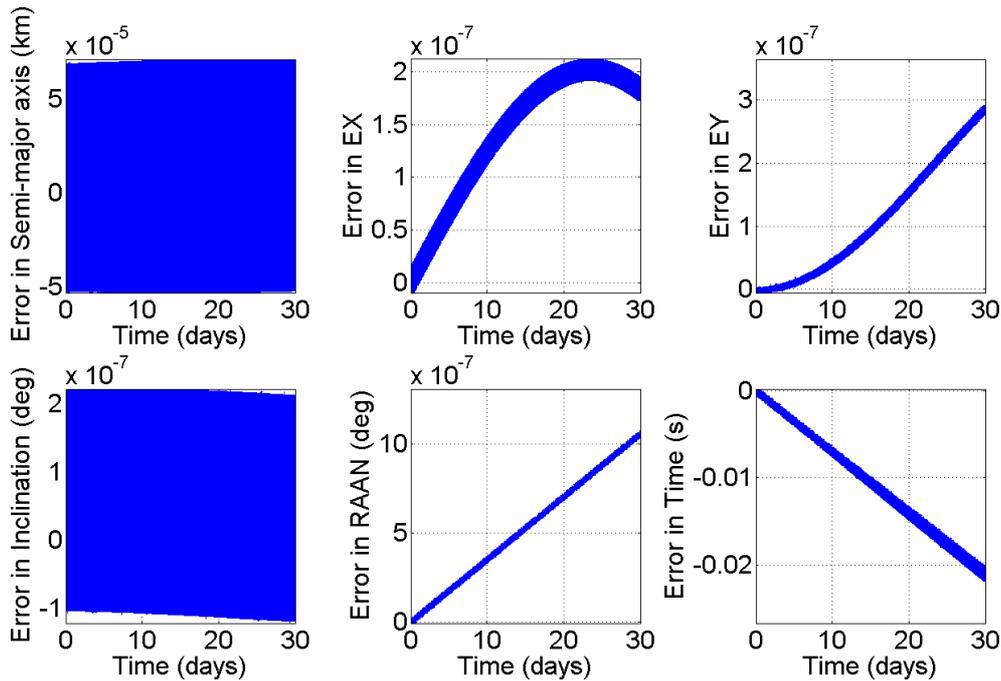}}
	\caption{Second order solution error in a long-term propagation of a near circular frozen orbit.}
	\label{fig:long_50deg}
\end{figure}

On the other hand, Fig.~\ref{fig:error_50deg} shows the error associated with the second order solution provided by this work. As can be seen, the error is small and comparable with a relative error of $J_2^2$ as expected from this kind of perturbation method. This error corresponds to a maximum error in position of 8.66 cm during the first orbital period. Moreover, it is also interesting to assess this error for more long-term propagations. To that end, Fig.~\ref{fig:long_50deg} shows the error for a 30-day propagation. This error corresponds to a maximum error in position of 7.5 meters after this propagation time. This shows the good long-term accuracy performance of the proposed approach.

\begin{figure}[h!]
	\centering
	{\includegraphics[width = 0.8\textwidth]{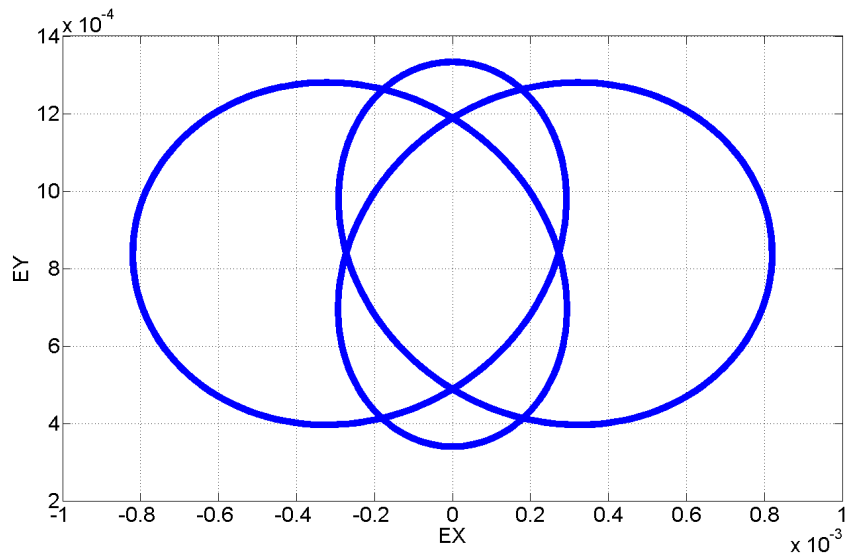}}
	\caption{Long-term evolution of the osculating value of the eccentricity for a near circular frozen orbit.}
	\label{fig:ecc_50deg}
\end{figure}

Additionally, it is also interesting to evaluate the evolution of the eccentricity vector over time. With that purpose in mind, Fig.~\ref{fig:ecc_50deg} shows the evolution of the eccentricity vector during a complete one-year propagation using a numerical propagator applied to the initial conditions that are defined with Eq.~\eqref{eq:frozen2}. As can be seen, even for a very long-term propagation, the eccentricity vector remains frozen.

\subsection{Eccentric frozen orbit close to the critical inclination}

As a second example of application, an orbit with a periapsis at 650 km of altitude and eccentricity of $e_{y0} = 0.2$ is selected. This means that the periapsis is located over the Northern hemisphere of the Earth. Additionally, we consider the initial condition $\theta_0 = 90$ deg and $e_{x0}$ to make the orbit symmetric with respect to a vertical plane passing trough the center of the Earth and the periapsis of the orbit. Note that, following the result from Section~\ref{sec:frozen_crit}, we have the possibility to use any value of $e_{x0}$ as long as $\mathcal{O}(e_{x0}) = \mathcal{O}(J_2)$. The case of modifying the initial condition $e_{x0}$ is covered later in this example. Finally, we set $\Omega_0 = 0$ deg for the right ascension of the ascending node.

\begin{figure}[h!]
	\centering
	{\includegraphics[width = 0.95\textwidth]{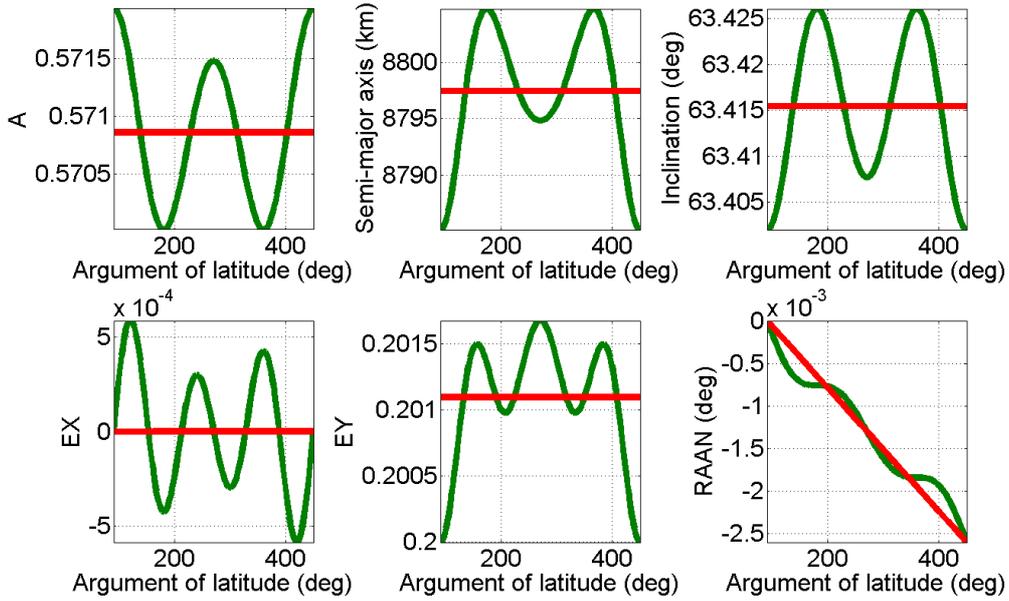}}
	\caption{Osculating and mean evolution of a near circular frozen orbit.}
	\label{fig:mean_ecc}
\end{figure}

These initial conditions lead to a resultant frozen orbit with an initial osculating inclination $i_0 = 63.402$ deg using Eqs.~\eqref{eq:kvalue} and~\eqref{eq:critexp}. Figure~\ref{fig:mean_ecc} shows the evolution of the orbital elements as well as the transformation from osculating to mean elements. As in the previous example, we can see that the semi-major axis, inclination and eccentricity vector remain periodic due to the frozen condition.

\begin{figure}[h!]
	\centering
	{\includegraphics[width = 0.95\textwidth]{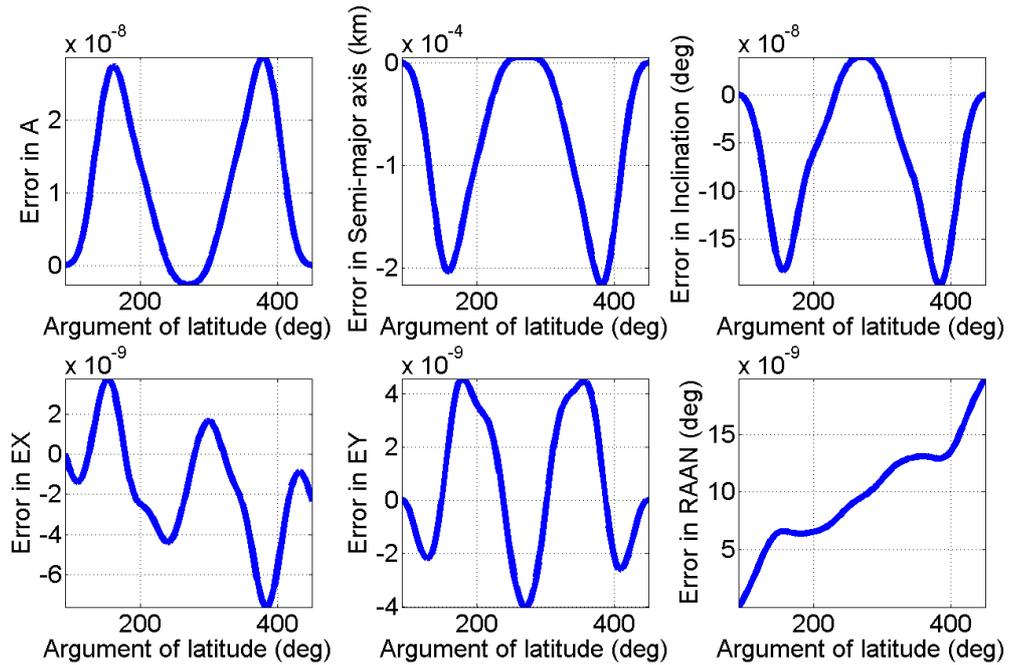}}
	\caption{Second order solution error of a near circular frozen orbit.}
	\label{fig:error_ecc}
\end{figure}

\begin{figure}[h!]
	\centering
	{\includegraphics[width = 0.95\textwidth]{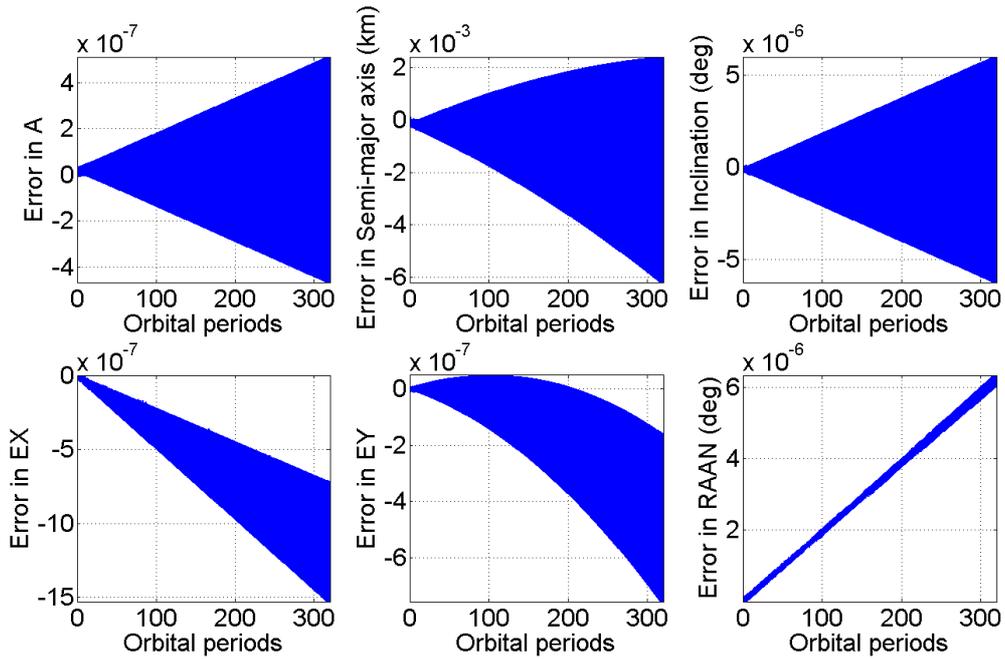}}
	\caption{Second order solution error in a long-term propagation of a near circular frozen orbit.}
	\label{fig:long_ecc}
\end{figure}

Additionally, it is possible to study the error associated with the second order analytical solution. Figure~\ref{fig:error_ecc} shows the error for an orbital period. This corresponds to a maximum error of less than 19 cm in position, and is coherent with the expected result from the perturbation method used. Moreover, Fig.~\ref{fig:long_ecc} presents the error of this solution when a propagation of 30 days is performed using the same initial conditions. As can be seen, the error increases over time as other analytical perturbation techniques, having a maximum of 54 m of error in position at the end of this propagation time.

\begin{figure}[h!]
	\centering
	{\includegraphics[width = 0.8\textwidth]{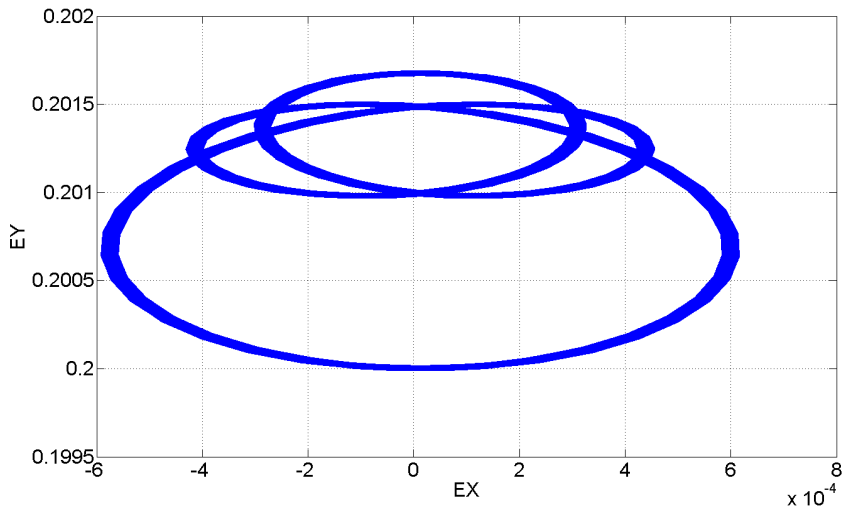}}
	\caption{Long-term evolution of the osculating value of the eccentricity for a near circular frozen orbit (symmetric orbit).}
	\label{fig:ecc_ecc}
\end{figure}

\begin{figure}[h!]
	\centering
	{\includegraphics[width = 0.8\textwidth]{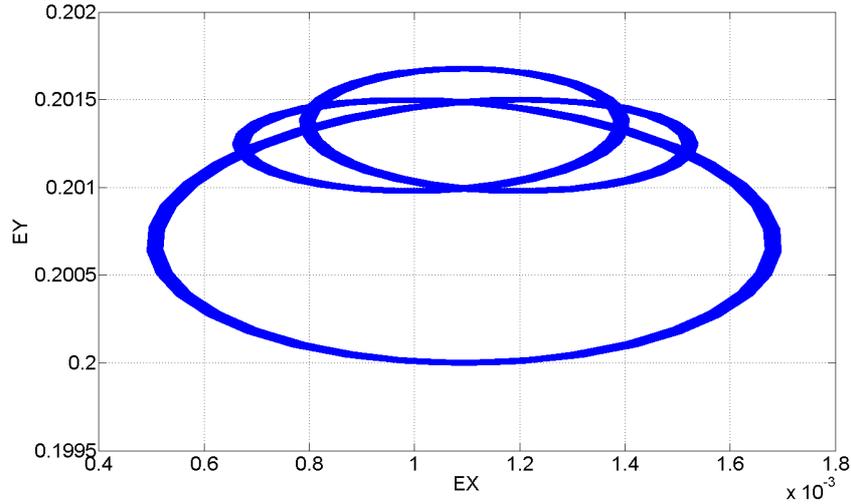}}
	\caption{Long-term evolution of the osculating value of the eccentricity for a near circular frozen orbit (non-symmetric orbit).}
	\label{fig:ecc_ecc2}
\end{figure}

Finally, Fig.~\ref{fig:ecc_ecc} presents the evolution of the eccentricity vector for a one-year propagation. As can be seen, the eccentricity vector remains frozen during this long-term propagation as in the previous example. Additionally, it is possible to study the effect of varying $e_{x0}$ on the eccentricity vector. To that end, Fig.~\ref{fig:ecc_ecc2} shows the evolution of the eccentricity vector when $e_{x0} = J_2$ while maintaining the other initial conditions and propagation time the same as before. This figure clearly shows the effect that the second order solution was predicting, the orbit maintains its argument of perigee and has low sensitivity to changes in $e_{x0}$.


\section{Conclusions}

This work presents a second order perturbation method to study the frozen orbits that appear in the zonal harmonics problem. Particularly, the zonal terms $J_2$, $J_3$, $J_4$, $J_5$, $J_6$, and $J_7$ are considered in this work. This is done using a power series expansion on the small parameter similar to the Poincar\'e-Lindstedt method but without the control in the perturbed frequency of the solution. This approach allows to have a good accuracy of the solution close to frozen orbits while reducing the size of the resultant expressions when compared with the Poincar\'e-Lindstedt method, therefore enabling an easier analysis of the system. Based on this approach, two different formulations are considered, the first one for studying near circular orbits, and the second one to study  more eccentric orbits.

The results of applying this perturbation approach are used to determine the osculating initial conditions that define a frozen orbit under a second order solution subject to the zonal harmonics terms considered. In that regard, several families of frozen orbits are analyzed and defined in closed-form in this work. This includes the near circular frozen orbits, as well as the families of frozen orbits that appear close to the critical inclination. An interesting result of this study is that the only condition for the $x$ component of the eccentricity vector in order to have a frozen orbit close to the critical inclination is that $\mathcal{O}(e_{x0}) = \mathcal{O}(J_2)$. This is an effect that was already identified and numerically tested for the $J_2$ problem, and it extends to the more general zonal harmonics problem. The numerical tests performed corroborate this finding.

Additionally, the methodology presented allows to study the bifurcation that appears in the frozen orbits close to the critical inclination. As a result of this, a closed-form expression is derived to find and evaluate these frozen orbits for any combination of initial conditions. These results show that the morphology of the bifurcation is highly dependent on the magnitude of angular momentum of the orbit.

Finally, the examples of application show that the proposed methodologies have a good accuracy performance even for long-term propagations. This includes the numerical propagation of the frozen orbits that are defined based on the initial conditions derived in this work.


\section*{Acknowledgments}

The author wants to dedicate this work to his grandparents Florencio Arnas Tello, Mar\'ia Escart\'in Mart\'in, Justa Ballesteros Bele\~no and Francisco Mart\'inez Tevar. I hope that one day I will be able to seen you again.



\end{document}